\documentclass[journal]{IEEEtran}

\usepackage[cmex10]{amsmath}
\usepackage{cite, amsfonts, amssymb, bm, bbm, graphicx, color, booktabs, multirow, relsize}
\usepackage[american]{babel}
\usepackage[T1]{fontenc}

\newtheorem{lemma}{Lemma}
\newtheorem{observation}{Observation}

\DeclareMathOperator*{\argmax}{arg\,max}
\DeclareMathOperator*{\tr}{Tr}
\DeclareMathOperator*{\diag}{diag}

\def \SINR{\text{\relsize{-2}\sf SINR}}

\def \CR{\text{\relsize{-2}\sf CR}}

\title{Joint Design of Overlaid Communication \\ Systems and Pulsed Radars}
\author{Le Zheng,~\IEEEmembership{Member,~IEEE,} Marco Lops,~\IEEEmembership{Senior Member,~IEEE,} \\ Xiaodong Wang,~\IEEEmembership{Fellow,~IEEE,} and Emanuele Grossi, ~\IEEEmembership{Senior Member,~IEEE}
\thanks{Le Zheng and Xiaodong Wang are with the Electrical Engineering Department, Columbia University, New York, USA, 10027, e-mail: le.zheng.cn@gmail.com, wangx@ee.columbia.edu. Marco Lops and Emanule Grossi are with the DIEI, Universita degli Studi di Cassino e del Lazio Meridionale, Cassino 03043, Italy (e-mail: lops@unicas.it, e.grossi@unicas.it).}}

\begin{document}

\maketitle

\begin{abstract}
The focus of this paper is on co-existence between a communication system and a pulsed radar sharing the same bandwidth. Based on the fact that the interference generated by the radar onto the communication receiver is intermittent and depends on the density of scattering objects (such as, e.g., targets), we first show that the communication system is equivalent to a set of independent parallel channels, whereby pre-coding on each channel can be introduced as a new degree of freedom. We introduce a new figure of merit, named the {\em compound rate}, which is a convex combination of rates with and without interference, to be optimized under constraints concerning the signal-to-interference-plus-noise ratio (including {\em signal-dependent} interference due to clutter) experienced by the radar and obviously the powers emitted by the two systems: the degrees of freedom are the radar waveform and the afore-mentioned encoding matrix for the communication symbols. We provide closed-form solutions for the optimum transmit policies for both systems under two basic models for the scattering produced by the radar onto the communication receiver, and account for possible correlation of the signal-independent fraction of the interference impinging on the radar. We also discuss the region of the achievable communication rates with and without interference. A thorough performance assessment shows the potentials and the limitations of the proposed co-existing architecture.
\end{abstract}


\section{Introduction}

Co-existence between radar and communication systems over overlapping (if not coincident) bandwidths has been a primary investigation field in recent years and has been put forward as a challenging topic at both theoretical and implementation stages~\cite{griffiths2014t09, griffiths2015radar, deng2013interference, hessar2016spectrum}. The prevailing approach so far---with some exception---has been to guarantee the detection and estimation performance of the radar by designing waveforms producing a tolerable level of interference on the communication system.

Some early results concerning the preservation of radar detection capabilities in the presence of co-existing---possibly un-licensed---wireless users have been established in~\cite{aubry2014radar, aubry2014cognitive, aubry2015new}: the approach taken here relies on maximizing the Signal-to-Interference-plus-Noise Ratio (SINR) in signal-dependent clutter under constraints concerning both the interference produced on overlaid communication networks and similarity with a standard waveform, possibly in the presence of some side information on the environment. A combined approach based on Mutual Information (MI) and SINR is developed in~\cite{turlapaty2014joint}. The philosophy of dual-function radar-communication relies instead on regarding the radar as the primary function and the communication as a secondary one, whose data can be embedded in the radar waveform~\cite{blunt2010embedding, hassanien2016signaling, hassanien2016dual, hassanien2016non}. A more Information-Theoretic approach is instead taken in~\cite{bica2016mutual}, wherein the object of interest is again the radar, dealt with through MI, under the constraint that it does not produce excessive interference on the coexisting communication system. The attention is steered back to the performance of the communication system in~\cite{lioptimum}, wherein Matrix-Completion based Multiple-Input Multiple-Output (MIMO) radars are made to co-exist with wireless systems by constraining the average capacity of the latter, while minimizing a measure of the interference induced on the former. Communications and Radar functions are likewise given equal weight in~\cite{chiriyath2016inner, Paul2016joint}, aimed at investigating the interplay between the estimation accuracy in target localization on the delay-Doppler plane and the performance of a multiple-access coexisting system, characterized through the rate achievability regions of the active users.

The present contribution is aimed at further investigating the achievable performance of spectrally overlapping radar and communication systems by conjugating the detection-based approach of~\cite{aubry2015new, aubry2014cognitive, aubry2014radar} and~\cite{hassanien2016signaling, hassanien2016dual, hassanien2016non} with the more recent trends safeguarding also the communication rate. To this end, we exploit the fact that, while the interference produced by the communication system onto the radar is {\em persistent}, the interference produced by the radar onto the communication link is {\em intermittent}, namely, is tied to the duty cycle of the transmitted waveforms and to the number of objects reflecting towards the communication system. On the other hand, while the radar system may rely on range-gating to cope with multiple target situations and is typically equipped with clutter-reduction devices, the communication system is completely unprotected against spurious reflections produced by the radar signal and hitting the communication receiver. Considering the scenario of a pulsed radar whose basic pulse is equal to that employed by the communication system (consistent with~\cite{lioptimum}), the contribution of this paper is two-fold. At the system model level, we show that, exploiting the range-gating induced by the radar onto the communication system, the latter can be viewed as an ensemble of parallel, independent channels, whose maximum rates bounce between that of a possibly fading additive white Gaussian noise channel, call it $R_0$, and that of an interference channel, call it $R_1$. At the design level, we set up the problem of determining the optimum radar waveform and communication system code-book as the solution of the constrained maximization of a convex combination of the above two rates, called compound rate, subject to the condition that the average SINR experienced by the radar---in turn affected by the interference from the communication system, by {\em signal-dependent interference} due to possible point-like clutters, and additive, possibly correlated signal-independent noise---exceeds a given minimum level. We offer closed-form solutions under two major situations of relevant theoretical and practical interest, that we name ``coherent'' and ``incoherent'' scattering: in the former, the objects producing scattering towards the communication receiver produce a delayed and randomly attenuated replica of the transmitted radar signal, as it happens, e.g., when they are stationary, while in the latter they produce a ``scintillating'' echo, as it happens, e.g., as they have random Doppler shift with uniform distribution. For the latter situation, we also offer an in-depth discussion on practical issues concerning the feasibility of the optimum solution. As a side result, we also derive the  region of the {\em achievable} rate pair $(R_0,R_1)$ in some relevant cases. A thorough performance assessment is also given to validate the proposed approach, showing that the nature of the scattering and the density of the interferers has a deep impact on the performance of the communication system, but  remarkable gains can be obtained through co-design especially if the requisites on the radar detection capabilities are particularly stringent.

The paper outline is as follows. Sec.~\ref{sec_signal_model} is devoted to the signal model and the problem formulation, while Sec.~\ref{sec:codesign} focuses on the constrained optimization of the aforementioned compound rate under white and colored radar noise. Sec.~\ref{sec_analysis} is devoted to the performance assessment and result validation, while conclusions and hints for future developments form the object of Sec.~\ref{sec_conclusion}.

\begin{figure}[t]
 \centering
%\centerline{\includegraphics[width=0.6\columnwidth]{fig_01.pdf}} % one-column version
 \centerline{\includegraphics[width=0.8\columnwidth]{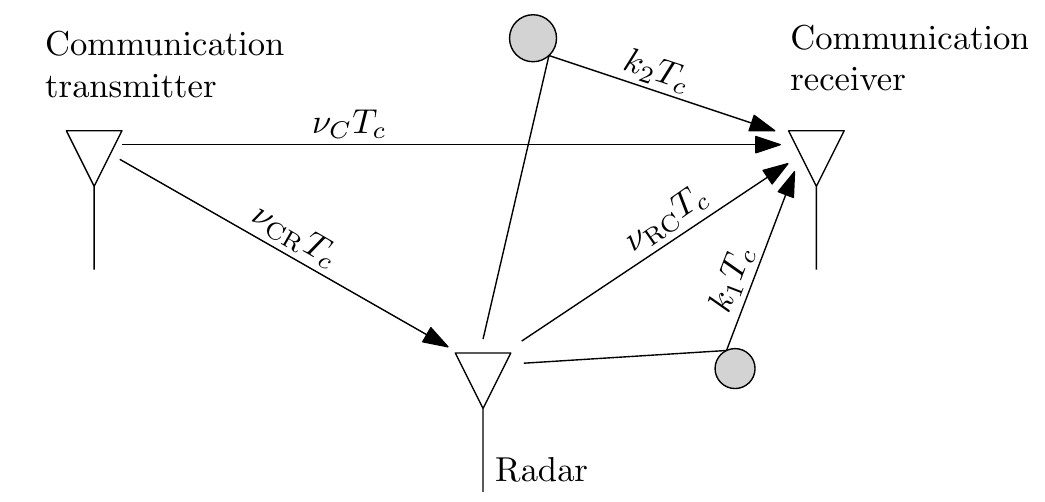}} % two-column version
 \caption{A possible scenario with two interference sources produced by the radar onto the communication system.} \label{fig_01}
\end{figure}

\section{System model and problem formulation}\label{sec_signal_model}

\begin{figure}[t]
 \centering
%\centerline{\includegraphics[width=0.95\columnwidth]{fig_02l.pdf}} % one-column version
\centerline{\includegraphics[width=1.01\columnwidth]{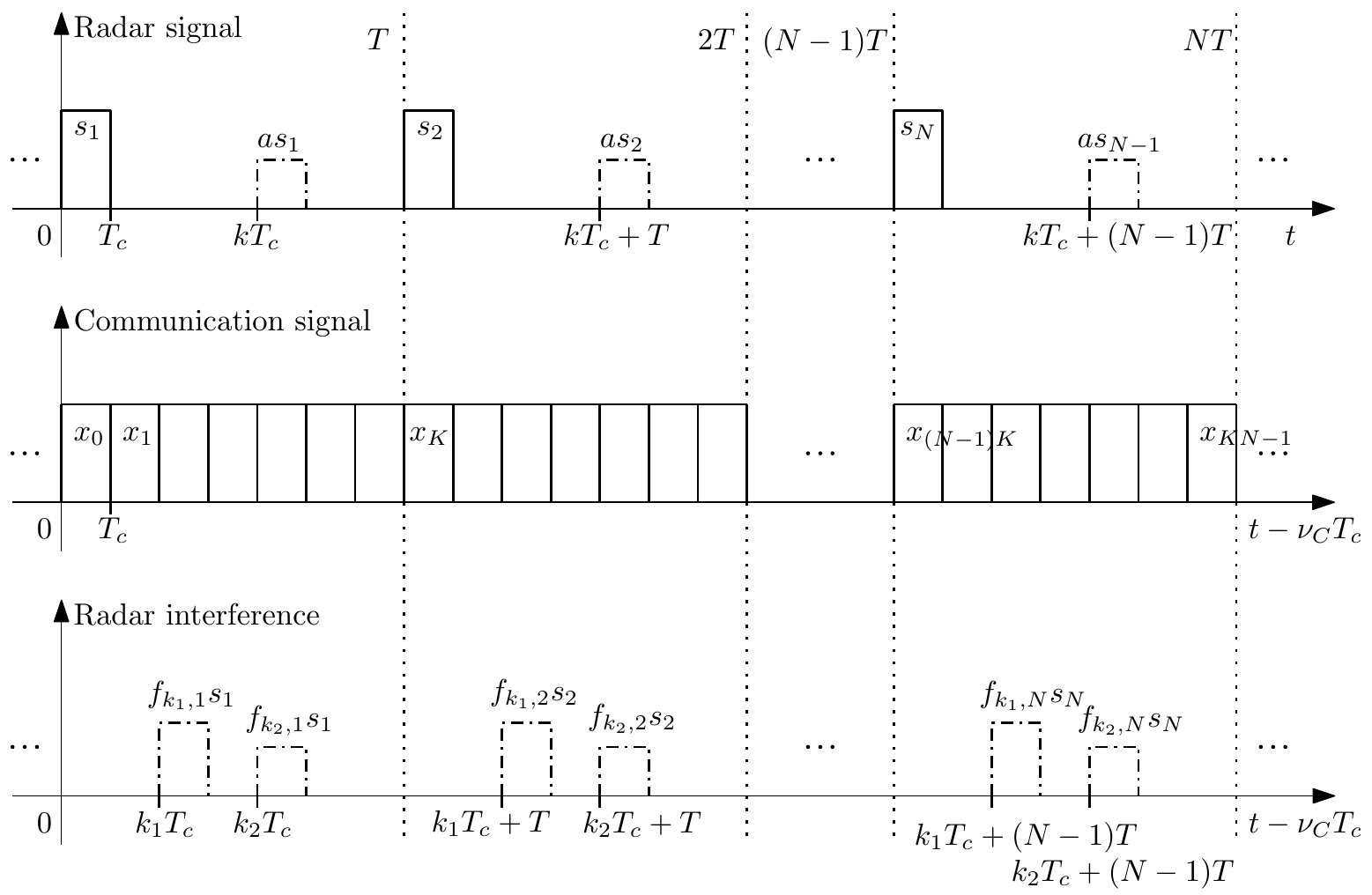}} % two-column version
 \caption{Example of the overlayed radar and communication. The first row is the transmitted radar signal and the echo from a target whose delay is $kT_c$. The superposition of the last two rows gives the received signal of communication.} \label{fig_02}
\end{figure}

In this paper, the joint radar-communication system consists of an active, mono-static, pulsed radar and a single-user communication system: Fig.~\ref{fig_01} outlines a scheme of the considered architecture. We assume that the radar and communication systems share the same bandwidth. The radar is allowed to transmit an amplitude-modulated pulse train with Pulse Repetition Time (PRT) $T$, each pulse having the same duration $T_c=T/K$, $K\in\mathbb{N}$, as the symbol duration in the communication system, so that the situation of the overlay is the one outlined in Fig.~\ref{fig_02}.

Suppose that the amplitude of the $n$-th radar pulse is $s_n$, so that the length-$N$ pulse train emitted by the radar is
\begin{equation}
 \xi(t) = \sum_{n=1}^N s_n \phi\big(t - (n-1)T\big)
\end{equation} 
where $\phi( \cdot )$ satisfies the Nyquist criterion with respect to $T_c$, whereby the bandwidth is $1/T_c$, and $K$ range cells are defined. As to $N$, we assume that it is chosen in such a way that no cell migration takes place in the time interval $NT$, i.e., that the targets do not change resolution cell for the duration of the pulse train. The signal transmitted by the communication system is
\begin{equation}
 \chi(t) = \sum_{\ell=-\infty}^{\infty} x_\ell \phi (t - \ell T_c)
\end{equation}
where $x_\ell$ denotes the $\ell$-th transmitted symbol.

If a target is present in the $k$-th range cell, $k\in\{1,\ldots,K\}$, it back-scatters towards the radar antenna with a given coupling coefficient $a$, and hits it with time delay $kT_c$, thus generating echoes at time instants $\left\{ kT_c + (n-1)T \right\}_{n=1}^N$: we assume that the targets of interest are coherent (e.g., follow a Swerling~0, I or III model), since no signal design could take place for scintillating targets (i.e., Swerling II or IV~\cite{Skolnik-2001}). Letting $\nu_{\text{CR}} T_c$ and $g$ be the delay and the channel gain between the communication transmitter and the radar receiver, respectively, the radar should process the return from the $k$-th range~\cite{Naghsh_2013} cell to discriminate between the two hypotheses
%\begin{equation}\label{radar_signal} % one-column version
% r(t) =\begin{cases}
% a \xi(t-kT_c) + g \chi(t - \nu_{\text{CR}}T_c) + c \xi(t-kT_c) + w(t), & \text{under } H_1 \\
% g \chi(t - \nu_{\text{CR}}T_c) + c \xi(t-kT_c) + w(t), & \text{under } H_0
%\end{cases}
%\end{equation}
\begin{equation}\label{radar_signal} % two-column version
 r(t) =\begin{cases}
 a \xi(t-kT_c) + g \chi(t - \nu_{\text{CR}}T_c) + c \xi(t-kT_c) + w(t),\\
 \hfill \text{under } H_1 \\
 g \chi(t - \nu_{\text{CR}}T_c) + c \xi(t-kT_c) + w(t),\hfill \text{under } H_0
\end{cases}
\end{equation}
where $w(t)$ is the measurement noise of the radar receiver, and $c$ is the scattering coefficient from all the other unwanted reflectors present (such as clutter and/or environmental reverberation) in the same range cell.\footnote{We have not accounted for the Doppler shift of the target, which boils down to considering it zero or to focusing on a specific Doppler resolution cell (e.g., the one with largest interference from the unwanted reflectors, in light of a worst-case design paradigm). Also, we are implicitly assuming point-like scatterers, which is quite reasonable in the narrowband low resolution scenario we are considering. This widely used signal model~\cite{Naghsh_2013} is simple but allows to capture the main performance tradeoffs.} The discrete-time representation of the observations pertaining to the $k$-th range cell (corresponding to a delay $kT_c$) after illumination through the pulse train can be obtained by projecting the previous observations onto the orthonormal functions $\left\{ \phi(t- k T_c-(n-1)T) \right\}_{n=1}^N$, i.e.,
\begin{align}
\label{eq:rn}
r_{k,n} &= \langle r(t), \phi\big(t - (n-1)T - k T_c \big) \rangle \notag \\
&=\begin{cases}
a s_n + g x_{k - \nu_{\text{CR}} +(n-1)K} + c s_n + w_{k,n}, & \text{under } H_1\\
g x_{k - \nu_{\text{CR}}+ (n-1)K} + cs_n +w_{k,n}, & \text{under }H_0
\end{cases}
\end{align}
where $\langle y(t), q(t) \rangle=\int_{-\infty}^\infty y(t) \operatorname{conj}\{q(t)\} dt$ is the inner product, $\operatorname{conj}\{\,\cdot\,\}$ denoting complex conjugation, and $w_{k,n} = \langle w(t), \phi\big(t  - k T_c- (n-1)T_c \big) \rangle$. Note that this dicretization process can be equivalently performed by standard filtering through a filter matched to the pulse waveform $\phi(t)$ and subsequent sampling at the inverse of the bandwidth $T_c$, as it is commonly done in radar receivers. Regrouping the returns pertaining to each range cell in $N$-dimensional vectors, we obtain, for the $k$-th range cell,
\begin{equation} \label{eq:radar-1}
\bm r_k = \begin{cases}
a \bm s + g \bm x_{k-\nu_{\text{CR}}} + c \bm s + \bm w_k, & \text{under }H_1\\
g \bm x_{k-\nu_{\text{CR}}} + c \bm s + \bm w_k, & \text{under } H_0
\end{cases}
\end{equation}
where $\bm r_k = [r_{k,1}\; r_{k,2}\; \cdots \; r_{k,N}]^T$, $\bm s = [s_1 \; s_2 \;\cdots \; s_N]^T$, $\bm x_m = [x_m\; x_{m+K}\; \cdots\; x_{m+ (N-1)K}]^T$, for $m$ integer, and $\bm w_k = [w_{k,1} \;w_{k,2} \; \cdots\; w_{k,N}]^T$, $(\,\cdot\,)^T$ denoting transpose. We assume that the distribution of $\bm w_k$ is ${\cal CN}({\bm 0}_N, \bm M)$, where $\bm 0_N $ is the all-zeros $N$-dimensional vector, $\bm M \in \mathbb{C}^{N \times N}$ is the covariance matrix, on whose diagonal elements is the unique value $\sigma^2_w$, representing the noise power, and ${\cal CN}$ denotes the complex circularly symmetric Gaussian distribution. Notice that here the time scale is the PRT, so the vector $\bm x_{k-\nu_\text{RC}}$ includes data spaced $T$ apart. This model implicitly assumes that the radar, although undertaking clutter reduction functions, may still be left with some signal-dependent interference, whose intensity is encapsulated in the mean square value of the random coefficient $c$; moreover, this clutter suppression functions might introduce some correlation between the noise samples, encapsulated in the matrix $\bm M$, which is assumed from now on known, e.g., as a consequence of accurate estimation conducted offline through secondary data.

As to the communication system, we denote by $h$ the gain of the channel linking the communication transmitter and radar receiver. In practice, $h$ can be obtained through the transmission of pilot signals~\cite{yin2013coordinated, simeone2004pilot}, so it is assumed known in this paper: it may be the realization of a random gain, thus yielding a block-fading channel, or a non-random constant. Let $\nu_{\text{C}} T_c$ be the delay between the transmitter and the receiver of the communication system. For sure, the communication receiver is not equipped, like the radar system, with interference and/or clutter suppression devices: the signal emitted by the radar produces echoes from a number of reflectors, each belonging to a given range-cell, resulting in  significant scattering towards the communication receiver (see Figs.~\ref{fig_01} and~\ref{fig_02}). In an overlay situation the timing information between the two systems is shared,\footnote{This boils down to assuming that frame synchronism is guaranteed, i.e., the communication system is made aware of the beginning of the train pulse. This is a low-rate information, which can be shared once and for all, and regularly updated to account for possible timing drifts~\cite{Scholtz1980, Ling2017}. For example, it can be reliably estimated offline, e.g., by using the direct path, or with a properly designed training phase.} whereby, letting $\nu_{\text{RC}}T_c$ be the delay from the radar transmitter to the communication receiver and $M$ be the total number of such interferers, the signal at the communication side can be cast as
\begin{align}
z(t) &= h \xi(t - \nu_{\text{C}} T_c ) + \sum_{m=1}^M \sum_{n=1}^N f_{k_m,n} s_n \notag\\
&\quad \times \phi\big(t - \nu_{\text{RC}}T_c - k_m T_c - (n-1) T\big) + v(t)
\end{align}
where both $M$ and $k_1,\ldots,k_M$ are unknown quantities,\footnote{Here we assume that the interferers' Doppler shift with respect to the communication system receiver is zero, i.e. that they are either stationary or in tangential motion. The subsequent results, however, hold true whenever the relevant figure of merit is Doppler-independent, as it happens for one of the two major situations considered in this paper, i.e., totally incoherent scattering, which has a particularly important meaning (see Sec.~\ref{sec:prob_formulation}). Once again, accounting for arbitrary interference Doppler shift would be compatible with the present framework, once a worst-case philosophy is embraced (see also~\cite{Naghsh_2014}).} $f_{k_m, n}$ is the reflection coefficient of the $m$-th interfering object\footnote{At the communication system side, we do not distinguish between target and clutter returns, for it is not concerned with target detection: any source of reflection causes interference and is therefore generally referred to as an interfering object.} in the $n$-th PRT, and $v(t)$ is the measurement noise of the communication receiver. As to the reflection coefficients $f_{k_m, n}$ they are typically random and will be commented upon later on in the paper. This model does not rule out, in principle, the case that one of the interferers is a direct path from the radar transmitter to the communication receiver, and the subsequent derivations would hold true in this case too. However, the most damaging sources of interference onto the communication system are those produced by objects---either targets or reverberation---whose angles and ranges are unknown, since the direct path might in principle be canceled through such techniques beam-forming~\cite{liu2014joint, chen2014adaptive}. By adopting the orthonormal basis $\{\phi(t-\nu_{\text{RC}}T_c-\ell T_c)\}_{\ell=-\infty}^\infty$, the $KN$ projections of the observations received in the interval $[\nu_{\text{RC}}T_c, \nu_{\text{RC}}T_c+NT]$ can be re-arranged in the form
%\begin{align} % one-column version
% z_{k,n} & = \langle z(t), \phi\big(t - \nu_{\text{RC}}T_c - kT_c - (n-1) T \big) \rangle \notag \\
% & = \begin{cases}
% h x_{k-\nu_{\text{C}}+\nu_{\text{RC}}+(n-1)K} + f_{k,n} s_n + v_{k,n}, &  \text{if } k \in \{k_1, \ldots , k_M\}  \\
% h x_{k-\nu_{\text{C}}+\nu_{\text{RC}}+(n-1)K} +  v_{k,n},  & \text{if } k \in \{1, \ldots , K\}\setminus \{k_1, \ldots , k_M\}.
% \end{cases}
%\end{align}
\begin{align} % two-column version
 z_{k,n} & = \langle z(t), \phi\big(t - \nu_{\text{RC}}T_c - kT_c - (n-1) T \big) \rangle \notag \\
 & = \begin{cases}
 h x_{k-\nu_{\text{C}}+\nu_{\text{RC}}+(n-1)K} + f_{k,n} s_n + v_{k,n},\\
 \hfill  \text{if } k \in \{k_1, \ldots , k_M\}  \\
 h x_{k-\nu_{\text{C}}+\nu_{\text{RC}}+(n-1)K} +  v_{k,n}, \\
 \hfill \text{if } k \in \{1, \ldots , K\}\setminus \{k_1, \ldots , k_M\}.
 \end{cases}
\end{align}
for $n=1, \ldots, N$, where $v_{k,n} = \langle v(t), \phi\big(t  - k T_c- (n-1)T_c \big) \rangle$. The previous equation highlights that co-existence induces a form of range gating on the communication system, whereby, regrouping the above observations into $K$ vectors of dimension $N$, whose entries represent samples, spaced $T$ apart, pertaining to a given range cell, we obtain the model for the communication signal:
\begin{equation} \label{eq:model-1}
 \bm z_k =\begin{cases}
 h \bm x_{k-\nu_{\text{C}}+\nu_{\text{RC}}} + \bm S \bm f_k + \bm v_k,  \text{ if } k \in \{k_1, \ldots , k_M\}  \\
 h \bm x_{k-\nu_{\text{C}}+\nu_{\text{RC}}} + \bm v_k, \text{ if } k \in \{1, \ldots , K\}\setminus \{k_1, \ldots , k_M\}
 \end{cases}
\end{equation}
where $\bm z_k = [z_{k,1}\; z_{k,2}\; \cdots \; z_{k,N}]^T$, $\bm S = \diag(s_1,s_2, \ldots,s_N) \in \mathbb{C}^{N \times N}$, $\diag(\,\cdot\,)$ denoting the diagonal matrix whose diagonal entries are the input elements, $\bm f_k = [f_{k,1}\; f_{k,2}\; \cdots \; f_{k,N}]^T$, and $\bm{v}_k = [v_{k,1} \; v_{k,2} \; \cdots \; v_{k,N}]^T$. Here we assume that $\bm{v}_k \sim {\cal CN}(\bm 0_N, \sigma_v^2 \bm I_N)$, with $\bm I_N$ denoting the $N$-dimensional identity matrix; also, we assume that\footnote{This assumption will be justified shortly.} $\bm f_k \sim {\cal CN}(\bm 0_N, \bm R_{f,k})$: in other words, the radar produces a signal-dependent interference onto the communication system, the nature of the reflecting object being encapsulated in the covariance matrix $\bm R_{f,k}$, on whose diagonal elements is a unique value, $\sigma^2_{f,k}$ say, representing the intensity of the scattered power.

\begin{figure*}[t] 
 \centering
% \centerline{\includegraphics[width=0.9\columnwidth]{fig_03.pdf}}% one-column version
  \centerline{\includegraphics[width=0.7\textwidth]{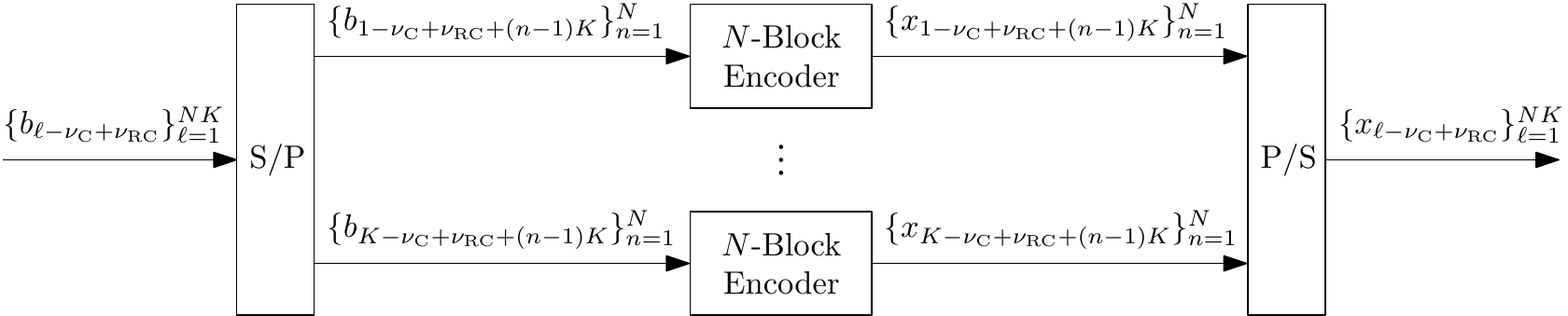}} % two-column version
 \caption{The information symbols $\{b_\ell\}$ are serial-to-parallel converted into $K$ streams, whose symbols are spaced $K$ epoch apart. Each stream is then encoded through an $N$-length block code with Gaussian codewords with correlation $\bm R_x$. The $K$ streams are then parallel-to-serial converted and transmitted.} \label{fig_03}
\end{figure*}
In principle, the communication system alone would achieve capacity by random coding through independent Gaussian codewords. In a co-existing architecture, a new degree of freedom is introduced, i.e., the covariance matrix $\bm R_x=\mathbb{E}\big[ \bm x_k \bm x_k^H \big] =\mathbb{E} \big[\bm x_{k-\nu_{\text{CR}}} \bm x_{k-\nu_{\text{CR}}}^H \big] =\mathbb{E}\big[ \bm x_{k-\nu_{\text{C}}+\nu_{\text{RC}}} \bm x_{k-\nu_{\text{C}}+\nu_{\text{RC}}}^H \big]$, $k = 1,\ldots,K$, where $(\,\cdot\,)^H$ denotes conjugate transpose. This form of ``rake encoding'' corresponds to introducing correlation between coordinates spaced $KT_c$ apart; on a different point of view, it amounts to generating a white $N \times K$-dimensional matrix, inducing the correlation $\bm R_x$ between the elements of each column, and undertaking depth-$K$ interleaving (i.e., transmitting the elements along the rows). Fig.~\ref{fig_03} shows an implementation of this encoding scheme. Also notice that the subscripts $\nu_{\text{CR}}$ in~\eqref{eq:radar-1} and $\nu_{\text{C}},\nu_{\text{RC}}$ in~\eqref{eq:model-1} become now irrelevant, whereby they will be omitted in subsequent derivations.

In next Section, we introduce the key performance measure that will be used for joint optimization of the communication and radar systems, while, in Sec.~\ref{sec:prob_formulation}, we present the optimization problem tackled in this paper.

\subsection{Performance measures}\label{sec:perf_measures}

For the radar system, the SINR is used as the figure of merit, which is expressed as
\begin{align}
 \SINR(\bm R_x, \bm s) & = \tr\left( (\sigma_g^2 \bm R_x + \sigma^2_c \bm s \bm s^H + \bm M)^{-1} \sigma_a^2 \bm s \bm s^H \right) \notag\\
 & = \sigma_a^2 \bm s^H (\sigma_g^2 \bm R_x  + \sigma^2_c \bm s \bm s^H + \bm M)^{-1} \bm s \label{SINR}
\end{align}
where $\sigma^2_a$, $\sigma^2_c$, and $\sigma^2_g$ are the variances of $a$, $c$ and $g$, respectively.\footnote{In what follows, the dependency of SINR on $(\bm R_x, \bm s)$ will be omitted whenever possible, so as to make the notation lighter.} The SINR is a key performance measure~\cite{blackman1999design, aubry2014radar, aubry2015new} and is also closely related to another pivotal figure of merit used for detection optimization purposes: the pair of Kullback-Leibler divergences between the densities of the observations under the two alternative hypotheses~\cite{Naghsh_2013, kay2009waveform}. Indeed, the two divergences can be interpreted, in the light of the Chernoff-Stein Lemma~\cite{cover2012elements}, as error exponents of miss and false alarm probabilities in a Neyman-Pearson theoretic environment, and, in the framework of sequential decision rules~\cite{poor2009quickest}, they offer guarantees on the average sample number necessary to make decisions. Also, the divergences have already been validated as valid alternatives to the usual probabilities of detection and false-alarm for waveform design purposes~\cite{grossi2012space}. For the observation model in~\eqref{eq:radar-1}, when $g$ is deterministic and $a\sim\mathcal{CN}(0,\sigma^2_g)$, maximizing the SINR is equivalent to maximizing the Kullback-Leibler divergences, as it is shown in Appendix~\ref{proof_KL_div}, which reinforces the choice of adopting the SINR as a figure of merit in our framework.

For the communication system, assuming that rake-type random coding with Gaussian codewords is used, the rate of the $k$-th channel bounces between that of an additive white Gaussian noise channel, i.e.,
\begin{equation}
\label{eq:C0}
R_0 (\bm R_x) = \frac{1}{N}\log \det \left( \bm I_{N} + \frac{|h|^2}{\sigma_v^2} \bm R_{x} \right) \quad \text{[bits/channel use]}
\end{equation}
and that of an interference channel, where the interference has covariance matrix $\sigma^2_v \bm I_N+\bm S \bm R_{f,k} \bm S^H$. In principle, such an interference might not be Gaussian (e.g., if the reflectors represent clutter); however, on top of the fact that we are considering low-resolution systems, wherein the Gaussian assumption might be justified,  at the design stage and inasmuch as the communication system performance is considered the Gaussian model has strong theoretical motivations (e.g., in the light of mini-max principle \cite[p. 298]{cover2012elements}), whereby we assume it outright, and the rate takes on the form
%\begin{equation} % one-column version
%R_{1,k} (\bm R_x, \bm s)=  \frac{1}{N} \log \det \left( \bm I_N + \frac{|h|^2}{\sigma _v^2} \bm R_x \left( \bm I_N} + \frac{1}{\sigma _v^2}\bm S \bm R_{f,k} \bm S^H \right)^{-1} \right).
%\end{equation}
\begin{align} % two-column version
R_{1,k} (\bm R_x, \bm s) &=  \frac{1}{N} \log \det \Bigg( \bm I_N + \frac{|h|^2}{\sigma _v^2} \bm R_x\notag \\
&\quad \times \left( \bm I_N + \frac{1}{\sigma _v^2}\bm S \bm R_{f,k} \bm S^H \right)^{-1} \Bigg).
\end{align}
Also, we let $\bm R_f$ be the ``worst case'' covariance\footnote{In this paper no form of cognition is assumed, whereby the basic philosophy must necessarily be that of the ``worst case.''} of the reflectors (more on this {\em infra}), and define $\sigma^2_f$ the common value of its diagonal elements; this implies that $R_{1,k}(\bm R_x, \bm s)\geq R_1(\bm R_x, \bm s)$, with
%\begin{equation}\label{eq:C1} % one-column version
% R_1 (\bm R_x, \bm s) = \frac{1}{N} \log \det \left( \bm I_N + \frac{|h|^2}{\sigma _v^2}{\bm R_x} \left( \bm I_N + \frac{1}{\sigma _v^2}\bm S \bm R_{f} \bm S^H \right)^{-1} \right).
%\end{equation}
\begin{align}\label{eq:C1} % two-column version
 R_1 (\bm R_x, \bm s) & = \frac{1}{N} \log \det \Bigg( \bm I_N + \frac{|h|^2}{\sigma _v^2}{\bm R_x} \notag\\
 & \quad \times \left( \bm I_N + \frac{1}{\sigma _v^2} \bm S \bm R_{f} \bm S^H \right)^{-1} \Bigg).
\end{align}
The objective function we propose is the convex combination of the rates in~\eqref{eq:C1} and~\eqref{eq:C0}, hereinafter referred to as {\em compound rate} (CR), defined as\footnote{In what follows, the dependency of CR on $(\bm R_x, \bm s)$ will be omitted whenever possible, so as to make the notation lighter.}
%\begin{align} % one-column version
%\CR (\bm R_x, \bm s) & = \beta R_1 + (1-\beta) R_0 \notag \\
%&= \frac{\beta}{N} \log \det \left( \bm I_N + \frac{|h|^2}{\sigma _v^2}{\bm R_x} \left( \bm I_N + \frac{1}{\sigma _v^2}\bm S \bm R_f \bm S^H \right)^{-1} \right) \notag\\
%&\quad + \frac{1 - \beta}{N} \log \det \left( \bm I_N + \frac{|h|^2}{\sigma _v^2} \bm R_x \right)\label{eq:CR_def}
%\end{align}
\begin{align} % two-column version
\CR (\bm R_x, \bm s) & = \beta R_1 + (1-\beta) R_0 \notag \\
&= \frac{\beta}{N} \log \det \Bigg( \bm I_N + \frac{|h|^2}{\sigma _v^2}{\bm R_x} \notag\\
&\quad \left. \times \left( \bm I_N + \frac{1}{\sigma _v^2}\bm S \bm R_f \bm S^H \right)^{-1} \right) \notag\\
&\quad + \frac{1 - \beta}{N} \log \det \left( \bm I_N + \frac{|h|^2}{\sigma _v^2} \bm R_x \right)\label{eq:CR_def}
\end{align}
where $\beta \in [0,1]$.

The above choice deserves some further comments. Denote $\zeta_k$ as the indicator of the presence ($\zeta_k=1$) or absence ($\zeta_k=0$) of radar interference on the $k$-th channel. If there is no ``preferential'' range where the reflectors are located, we may assume that  $\{ \zeta_k\}_{k=1}^K$ are independent and identically distributed, with $\Pr( \zeta_k=1)=\alpha$. Thus, in the considered scenario, the presence of a co-existing radar system is accounted for by modeling the communication system as $K$ mutually independent parallel channels, each of them being interfered with probability $\alpha$. In this framework, when $\beta = \alpha$ and $\bm R_{f,k}=\bm R_f$ $\forall k$, the quantity $N \CR$ can be interpreted as the conditional MI, given the channel state, between the input and the output of the $k$-th channel,\footnote{In principle the input would be $\bm x_{k-\nu_{\text{C}}+\nu_{\text{RC}}}$, but since $I(\bm x_{k-\nu_{\text{C}}+\nu_{\text{RC}}}; \bm z_k\mid \zeta_k)=I(\bm x_k; \bm z_k\mid \zeta_k)$, with a slight notational abuse, we use here $\bm x_k$ as channel input.} i.e., $I(\bm x_k; \bm z_k\mid \zeta_k)=N\CR$; moreover $N\CR$ differs from the input-output MI, $I(\bm x_k; \bm z_k)$, by less than $1/N$. Consider indeed the identity
\begin{align}
 I(\bm x_k,\zeta_k;\bm z_k)&=I(\bm x_k ;\bm z_k)+I(\zeta_k;\bm z_k |\bm x_k)\notag\\
 & =I(\zeta_k;\bm z_k)+I(\bm x_k;\bm z_k|\zeta_k).
\end{align}
Since
\begin{align}
 I(\bm x_k ;\bm z_k |\zeta_k)&=\alpha I(\bm x_k ;\bm z_k |\zeta_k=1)\notag\\
 &\quad +(1-\alpha)I(\bm x_k ;\bm z_k |\zeta_k=0)\notag\\
 &=N\CR
\end{align}
the CR is in fact a conditional MI. Furthermore, we have
\begin{align}
 I(\bm x_k& ;\bm z_k)-N \CR\notag\\
 & = I(\zeta_k ;\bm z_k)-I(\zeta_k ;\bm z_k |\bm x_k), \notag \\
 &= H(\zeta_k)-H(\zeta_k|\bm z_k)-\big( H(\zeta_k|\bm x_k) - H(\zeta_k|\bm x_k,\bm z_k) \big), \notag \\
 &= -H(\zeta_k|\bm z_k)+H(\zeta_k|\bm x_k, \bm z_k)\notag\\
 & = -I(\zeta_k;\bm x_k |\bm z_k)
\end{align}
where $H(\,\cdot\,)$ denotes entropy and the independence between $\zeta_k$ and $\bm x_k$ has been exploited. As a consequence $\CR$ is {\em not} achievable through the proposed encoding scheme. However, since $0\leq I(\zeta_k;\bm x_k | \bm z_k) \leq 1$, we have that
\begin{equation} \label{eq:bound_mutual}
  \CR-\frac{1}{N}\leq  \frac{1}{N} I(\bm x_k ;\bm z_k)\leq \CR
\end{equation}
which shows that the MI per channel use differs from the CR by less than $1/N$. We hasten to underline here that, contrary to $\CR$, $\frac{1}{N} I(\bm x_k ;\bm z_k)$ represents the maximum {\em achievable} transmission rate, provided the pulse number $N$ is large enough. Notice that, lacking prior information on $\alpha$, the chosen value of $\beta$ (for optimization purposes) should depend on a coarse information (or forecast) on how crowded the scene is.

\subsection{Problem Formulation}\label{sec:prob_formulation}

The degrees of freedom available for optimization are the covariance matrix of the communication system $\bm R_x$ and the radar waveform $\bm s$. The objective function is the CR in~\eqref{eq:CR_def}, while a constraint is imposed on the minimum required SINR at the radar receiver~\cite{blackman1999design}, denoted $\rho_{\min}$, and on the maximum powers of the radar and of the communication system, denoted $P_r$ and $P_c$, respectively. Concerning $\rho_\text{min}$, it is usually set considering a reference target, i.e., a target set at a given distance (typically, the one corresponding to the $K$-th range cell) and following a given fluctuation model with a given average Radar Cross-Section (RCS)~\cite{Skolnik-2001}. The joint radar and communication waveform optimization problem could thus be set up in the following form:
\begin{equation}\label{eq:P0}
\begin{aligned}
\max_{\bm R_x, \bm s} &\quad \CR(\bm R_x, \bm s) \\
\text{s.t.} &\quad  \SINR(\bm R_x, \bm s) \geq \rho_\text{min}, \quad  \frac{1}{N}\|\bm s\|^2 \leq P_r\\
&  \quad \frac{1}{N}\tr(\bm R_x) \leq P_c, \quad \bm R_x \succeq 0
\end{aligned}
\end{equation}
where $\succeq$ denotes positive semi-definiteness.

Before presenting the solution to this problem, it is worthwhile giving the following comments.

\begin{observation}
The system model considered here relies on the assumption that the radar and the communication system sharing the same bandwidth are narrow-band: this implies that the channel is flat-fading, and that each reflector---whether it is a target the radar wants to detect or an interferer---scattering the radar signal towards the communication receiver produces a single resolvable path. Notice, however, that, should this not be the case, the mathematical setup of the design problem would require only updating the definition of the CR for the communication system. Indeed, an interferer producing resolvable paths would spread its reverberation across a number of different range cells (corresponding to delays that are multiple of $T_c$) with independent scattering coefficients\cite{Proakis_2001}. As a consequence, from the point of view of the communication system, this would translate into a denser environment (i.e., larger values of $\alpha$), while having no effect on the SINR, which is in fact evaluated focusing on a single range cell.
\end{observation}

\begin{observation}
The complex vector $\bm s$ in~\eqref{eq:radar-1} and~\eqref{eq:model-1} represents the slow-time code sequence of the radar, in that the coefficients $\{s_n\}$ encode pulses spaced PRT apart. However, a similar discrete-time data model can be set up for fast-time coding. In that case, the transmitted pulse has duration $NT_c$ and is composed of $N$ sub-pulses with bandwidth $1/T_c$; the coefficient $s_n$, instead, represents the amplitude of the $n$-th sub-pulse, as in~\cite{aubry2014radar, aubry2015new}; similarly, the codewords at the communication systems contain $N$ symbols spaced $T_c$ apart. However, the main advantage of designing the radar slow-time code is that we are not concerned with the stringent requirements---in terms of range resolution, peak-to-sidelobe levels of the correlation function, and, more generally, ambiguity function~\cite{Levanon_2004}---posed by fast-time coding. We underline here that the model lends itself to account for joint fast-time/slow-time coding, wherein a train of $N$ sophisticated pulses, each composed of $M$ encoded sub-pulses, is amplitude-modulated. The discrete-time model would be similar, with $\bm s$ having a Kronecker product structure with $MN$ entries, and the interference density $\alpha$ at the communication system being $M$-times larger. This problem is more challenging, since the constraints on the fast-time code must be included, and will be the subject of our future work. 
\end{observation}

\section{Waveform optimization} \label{sec:codesign}

Determining a general and closed-form solution to~\eqref{eq:P0} for arbitrary $\bm R_f$ appears unwieldy, but a deep insight into the consequences of having the two systems co-exist can be given by considering two important limiting situations, i.e.,
\begin{enumerate}\renewcommand{\theenumi}{\alph{enumi}}
 \item {\em Coherent} interference, such as, e.g., coherent targets yielding the rank-1 matrix $\bm R_f=\sigma^2_f \bm 1_N$, $\bm 1_N$ denoting the all-ones $N\times N$ matrix;
 \item {\em Incoherent} interference, such as scintillating scattering objects with low coherence time, yielding $\bm R_f=\sigma^2_f \bm I_N$.
\end{enumerate}
Here we consider, for both situations above, the case that the noise impinging on the radar is white. The case of colored noise is handled in Sec.~\ref{sec:incoher_design}, where the relationship between these two cases is also inspected. Finally, in Sec.~\ref{achievable_region}, the scope is enlarged beyond the optimization problem in~\eqref{eq:P0} by focusing the attention on the regions of achievable communication rate pairs.

\subsection{Coherent Interference} \label{sec:coher_design}

When $\bm R_f = \sigma_f^2 \bm 1_N$, the communication rate in~\eqref{eq:C1} can be rewritten as
%\begin{align} % one-column version
% R_1 (\bm R_x, \bm s) &= \frac{1}{N}\log \det \left( \bm I_N + \frac{|h|^2}{\sigma _v^2}{\bm R_x} \left( \bm I_N + \frac{\sigma _f^2}{\sigma _v^2} \bm s \bm s^H \right)^{-1} \right) \notag  \\
% &= \frac{1}{N}\left[\log \det \left( \bm I_N + \frac{|h|^2}{\sigma _v^2} \bm R_x \right) + \log \det \left( \bm I_N - \left( \bm I_N + \frac{|h|^2}{\sigma _v^2} \bm R_x \right)^{-1} \frac{|h|^2}{\sigma _v^2} \bm R_x \frac{\frac{\sigma _f^2}{\sigma _v^2} \bm s \bm s^H}{1 + \frac{\sigma _f^2}{\sigma _v^2} \| \bm s\|^2} \right)\right]  \notag \\
% &= \frac{1}{N} \left[\log \det \left( \bm I_N + \frac{|h|^2}{\sigma _v^2} \bm R_x \right) - \log \left( \frac{1 + \frac{\sigma _f^2}{\sigma _v^2} \| \bm s \|^2}{1 + \frac{\sigma _f^2}{\sigma _v^2} \bm s^H \left( \bm I_N + \frac{|h|^2}{\sigma _v^2} \bm R_x \right)^{-1} \bm s} \right)\right] \label{eq:R1rank1}
%\end{align}
\begin{align} % two-column version
 R_1 (\bm R_x, \bm s) &= \frac{1}{N}\log \det \left( \bm I_N + \frac{|h|^2}{\sigma _v^2}{\bm R_x} \left( \bm I_N + \frac{\sigma _f^2}{\sigma _v^2} \bm s \bm s^H \right)^{-1} \right) \notag  \\
 &= \frac{1}{N} \Bigg[ \log \det \left( \bm I_N + \frac{|h|^2}{\sigma _v^2} \bm R_x \right) + \log \det \Bigg( \bm I_N \notag\\
 &\quad \left. \left. - \left( \bm I_N + \frac{|h|^2}{\sigma _v^2} \bm R_x \right)^{-1} \frac{|h|^2}{\sigma _v^2} \bm R_x \frac{\frac{\sigma _f^2}{\sigma _v^2} \bm s \bm s^H}{1 + \frac{\sigma _f^2}{\sigma _v^2} \| \bm s\|^2} \right)\right]  \notag \\
 &= \frac{1}{N} \Bigg[\log \det \left( \bm I_N + \frac{|h|^2}{\sigma _v^2} \bm R_x \right) \notag\\
 &\quad\left.  - \log \left( \frac{1 + \frac{\sigma _f^2}{\sigma _v^2} \| \bm s \|^2}{1 + \frac{\sigma _f^2}{\sigma _v^2} \bm s^H \left( \bm I_N + \frac{|h|^2}{\sigma _v^2} \bm R_x \right)^{-1} \bm s} \right)\right] \label{eq:R1rank1}
\end{align}
where the last equality follows from the fact that $\det (\bm I_N + \bm p \bm q^H ) = 1 + \bm p^H \bm q$, with $\bm p, \bm q \in \mathbb{C}^N$, whereby, plugging~\eqref{eq:R1rank1} into~\eqref{eq:CR_def} and~\eqref{eq:P0}, the optimization problem can be reformulated as
%\begin{equation} one-column version
%\label{opt_prob_coher}
%\begin{aligned}
% \max_{\bm s, \bm R_x} & \quad \left\{ \log \det \left( \bm I_N + \frac{|h|^2}{\sigma_v^2}{\bm R_x} \right) - \beta \log \left( \frac{1 + \frac{\sigma _f^2}{\sigma _v^2} \| \bm s\|^2}{1 + \frac{\sigma _f^2}{\sigma _v^2} \bm s^H \left( \bm I_N + \frac{|h|^2}{\sigma _v^2}{\bm R_x} \right)^{-1} \bm s} \right) \right\} \\
% \text{s.t.} & \quad \sigma^2_a\bm s^H(\sigma_g^2 \bm R_x + \sigma^2_c\bm s \bm s^H + \sigma^2_w \bm I_N)^{-1} \bm s \geq \rho_\text{min}\\
%& \quad \frac{1}{N}\|\bm s\|^2 \leq P_r,  \quad \frac{1}{N}\tr(\bm R_x) \leq P_c, \quad \bm R_x \succeq 0.
%\end{aligned}
%\end{equation}
\begin{equation} % two-column version
\label{opt_prob_coher}
\begin{aligned}
 \max_{\bm s, \bm R_x} & \quad \Bigg\{ \log \det \left( \bm I_N + \frac{|h|^2}{\sigma_v^2}{\bm R_x} \right) \\
 &\quad \left. - \beta \log \left( \frac{1 + \frac{\sigma _f^2}{\sigma _v^2} \| \bm s\|^2}{1 + \frac{\sigma _f^2}{\sigma _v^2} \bm s^H \left( \bm I_N + \frac{|h|^2}{\sigma _v^2}{\bm R_x} \right)^{-1} \bm s} \right) \right\} \\
 \text{s.t.} & \quad \sigma^2_a\bm s^H(\sigma_g^2 \bm R_x + \sigma^2_c\bm s \bm s^H + \sigma^2_w \bm I_N)^{-1} \bm s \geq \rho_\text{min}\\
& \quad \frac{1}{N}\|\bm s\|^2 \leq P_r,  \quad \frac{1}{N}\tr(\bm R_x) \leq P_c, \quad \bm R_x \succeq 0.
\end{aligned}
\end{equation}
For the sake of completeness, here we also discuss the situation when $\bm s$ is optimized under given $\bm R_x$ and $\bm R_x$ is optimized under fixed $\bm s$, on the understanding that the main result is the joint design. The proofs can be found in Appendix~\ref{proof_coherent}.

\subsubsection{Fixed communication codebook}
In this case, the communication $\bm R_x$ is given,\footnote{This situation has a purely theoretical interest and is intended to show the difference between joint and single optimization, since in fact a non-coexisting communication system should use $\bm R_x  \propto \bm I_N$.} with $\frac{1}{N}\tr(\bm R_x)\leq P_c$. The radar system is overlaid, and its waveform must be properly designed by solving~\eqref{opt_prob_coher} with fixed $\bm R_x$. This problem admits a solution only if
\begin{equation}\label{existence_coher_fixed_R}
 \rho_\text{min} \leq \frac{\sigma^2_aN P_r}{\sigma^2_g \gamma_N +\sigma^2_cNP_r+ \sigma^2_w}
\end{equation}
where $\gamma_N$ is the smallest eigenvalue of $\bm R_x$, in which case, the optimal radar waveform is
\begin{equation} \label{opt_s_coher_fixed_R}
 \bm s^* = \sqrt{\frac{\rho_\text{min}(\sigma^2_g \gamma_N +\sigma^2_w)}{\sigma^2_a-\sigma^2_c\rho_\text{min}}} \; \bm u_N
\end{equation}
where $\bm u_N$ is the eigenvector of $\bm R_x$ corresponding to $\gamma_N$. This situation matches the intuition that the radar transmits, with minimum power compatible with the constraint, in the least-interfered direction of the signal space where.

\subsubsection{Fixed radar waveform}
In this case, the radar waveform $\bm s$ is given, with
\begin{equation}
\begin{cases}
 \frac{1}{N} \| \bm s \|^2 \leq P_r\\
 \rho_\text{min} \leq \frac{\sigma^2_a \| \bm s \|^2}{\sigma^2_c \| \bm s \|^2 +\sigma^2_w}
\end{cases}
\end{equation}
so that the radar constraints are satisfied when no interfering system is present. Once a communication system is overlaid, the covariance matrix of its codewords should solve~\eqref{opt_prob_coher} with $\bm s$ fixed. The solution is
\begin{equation} \label{opt_R_coher_fixed_s}
 \bm R_x^* = \bm U^*  \diag \left( \frac{NP_c - \gamma_N^*}{N-1} , \ldots ,  \frac{NP_c - \gamma_N^*}{N-1}, \gamma_N^* \right)  (\bm U^*)^H
\end{equation}
with $\bm U^*\in\mathbb{C}^{N\times N}$ any unitary matrix whose last column is $\frac{1}{\|\bm s\|}\bm s$, while the expression of $\gamma^*_N$ is reported in~Appendix~\ref{proof_coherent}, Eq.~\eqref{eq:optgammaN}. This strategy boils down to splitting the power between the $N-1$ interference-free eigenvectors---where noise whiteness explains the uniform allocation---and the direction of the radar signal. The fraction of power allocated to the interfered direction is dictated by $\gamma^*_N$, which depends on the system parameters and constraints.

\subsubsection{Joint optimization}
The problem is formulated by~\eqref{opt_prob_coher}, and admits a solution only if
\begin{equation}
 \rho_\text{min} \leq \frac{\sigma^2_a NP_r}{\sigma^2_c N P_r +\sigma^2_w}.\label{existence_coher_joint}
\end{equation}
In this case, the optimal covariance of the communication system is
\begin{equation}\label{opt_R_coher_joint}
 \bm R_x^* = \bm U^*  \diag \left( \frac{NP_c - \gamma_N^*}{N-1} , \ldots ,  \frac{NP_c - \gamma_N^*}{N-1}, \gamma_N^* \right)  (\bm U^*)^H
\end{equation}
where $\bm U^*\in\mathbb{C}^{N\times N}$ is any unitary matrix and $\gamma^*_N$ is reported in Eq.~\eqref{opt_gamma_N_coher_joint} of Appendix~\ref{proof_coherent}, while the optimal radar waveform is
\begin{equation}\label{opt_s_coher_joint}
 \bm s^* = \sqrt{ \frac{\rho_\text{min}(\sigma^2_g\gamma_N^*+\sigma^2_w)}{\sigma^2_a-\sigma^2_c\rho_\text{min}}} \bm u_N^*
\end{equation}
with $\bm u^*_N$ denoting the last column of $\bm U^*$. The previous equation clearly shows that, under joint optimization, the structure of the solution in~\eqref{opt_R_coher_joint} and~\eqref{opt_s_coher_joint} is similar to~\eqref{opt_R_coher_fixed_s} and~\eqref{opt_s_coher_fixed_R}.

\subsection{Incoherent interference}\label{sec:incoher_design}

When $\bm R_f = \sigma_f^2 \bm I_N$, the communication rate in~\eqref{eq:C1} can be rewritten as
%\begin{align} % one-column version
%R_1 (\bm R_x, \bm s) &= \frac{1}{N} \log \det \left( \bm I_N + |h|^2 \bm R_x \left( \sigma _v^2 \bm I_N + \sigma _f^2 \bm S \bm S^H \right)^{-1} \right) \notag\\
%&= \frac{1}{N}\log \det \left( \bm I_N + \frac{|h|^2}{\sigma^2_v} \bm R_x \diag \left( \left\{ \left(1 + \tfrac{\sigma_f^2}{\sigma _v^2} | s_i|^2\right)^{-1} \right\}_{i=1}^N \right) \right) \label{eq:R1incoherent}
%\end{align}
\begin{align} % two-column version
R_1 (\bm R_x, \bm s) &= \frac{1}{N} \log \det \left( \bm I_N + |h|^2 \bm R_x \left( \sigma _v^2 \bm I_N + \sigma _f^2 \bm S \bm S^H \right)^{-1} \right) \notag\\
&= \frac{1}{N}\log \det \Bigg( \bm I_N + \frac{|h|^2}{\sigma^2_v} \bm R_x \notag\\
& \quad \left. \times \diag \left( \left\{ \left(1 + \tfrac{\sigma_f^2}{\sigma _v^2} | s_i|^2\right)^{-1} \right\}_{i=1}^N \right) \right) \label{eq:R1incoherent}
\end{align}
whereby, plugging~\eqref{eq:R1incoherent} into~\eqref{eq:CR_def} and~\eqref{eq:P0}, the problem becomes
\begin{equation}\label{opt_prob_incoher}
 \begin{aligned}
 \max_{\bm R_x, \bm s} & \quad \Bigg\{ (1-\beta) \log \det \left( \bm I_N + \frac{|h|^2}{\sigma_v^2} \bm R_x \right) + \beta \log \det \Bigg( \bm I_N \\
 &\quad \left. + \frac{|h|^2}{\sigma^2_v} \bm R_x \diag \left( \left\{ \left(1 + \tfrac{\sigma_f^2}{\sigma _v^2} | s_i|^2\right)^{-1} \right\}_{i=1}^N \right) \right)  \Bigg\}\\
 \text{s.t.} & \quad \sigma^2_a \bm s^H(\sigma_g^2 \bm R_x + \sigma^2_c \bm s \bm s^H + \sigma^2_w\bm I_N)^{-1} \bm s \geq \rho_\text{min}\\
 & \quad \frac{1}{N} \|\bm s\|^2 \leq P_r, \quad \frac{1}{N}\tr(\bm R_x) \leq P_c, \quad \bm R_x \succeq 0.
 \end{aligned}
\end{equation}
This situation represents the case that the communication system is affected by scintillating interferers. Also, as shown in~\cite{Naghsh_2014}, it can model the case where, lacking any prior information as to the Doppler frequencies of the objects producing interference on the communication receiver, the Doppler shifts---normalized to $T$---are modeled as uniformly distributed on an interval of amplitude 1. 

As shown in Appendix~\ref{proof_incoherent}, Problem~\eqref{opt_prob_incoher} admits a solution only if
\begin{equation}
 \rho_\text{min} \leq \frac{\sigma^2_a NP_r}{\sigma^2_c N P_r +\sigma^2_w} \label{existence_incoher_joint}
\end{equation}
in which case, the optimal covariance matrix of the communication system and radar waveform are
\begin{subequations} \label{opt_R_s_incoher_joint}
\begin{align}
 \bm R_x^* & =  \diag \left( \frac{NP_c - \gamma_N^*}{N-1} , \ldots ,  \frac{NP_c - \gamma_N^*}{N-1}, \gamma_N^* \right) \\
 \bm s^* &= \sqrt{ \frac{\rho_\text{min}(\sigma^2_g\gamma_N^*+\sigma^2_w)}{\sigma^2_a-\sigma^2_c\rho_\text{min}}} [0 \; \cdots \; 0 \; 1]^T \label{opt_s_incoher_joint}
 \end{align}%
\end{subequations}
respectively, where $\gamma^*_N$ is reported in Eq.~\eqref{opt_gamma_N_incoher_joint} of Appendix~\ref{proof_incoherent}.

This solution is similar to the one for the coherent scattering in~\eqref{opt_s_coher_joint} and~\eqref{opt_R_coher_joint}, and results in the same optimized CR. However, the degree of freedom in the choice of the eigenvector matrix $\bm U^*$ is now lost, and this may lead to practical implementation problems. Indeed, the radar waveform in~\eqref{opt_s_incoher_joint} might not be feasible, since all of the energy is concentrated in a single pulse, and this can break practical limitations on the Peak-to-Average Power Ratio (PAPR). In this case, one could include in the design the additional constraint
\begin{equation}
 \max_{n\in\{1,\ldots,N\}} |s_n|^2 \leq \delta \frac{1}{N} \|\bm s\|^2
\end{equation}
where $\delta\in[1,N]$ is the maximum allowed PAPR. Unfortunately, no closed-form solution to this optimization problem, appears to be available. Nevertheless, one can always approximate~\eqref{opt_R_s_incoher_joint} as 
\begin{subequations}\label{naif_solution}
\begin{align}
 \bm R_x^* & = \bm U \diag \left( \frac{NP_c - \gamma_N^*}{N-1} , \ldots ,  \frac{NP_c - \gamma_N^*}{N-1}, \gamma_N^* \right) \bm U^H\\
 \bm s^* &= \sqrt{ \frac{\rho_\text{min}(\sigma^2_g\gamma_N^*+\sigma^2_w)}{\sigma^2_a-\sigma^2_c\rho_\text{min}}} \bm u
 \end{align}%
\end{subequations}
where
\begin{equation}
 \bm u= \left[ \sqrt{\frac{N-\delta}{N(N-1)}}  \; \cdots \; \sqrt{\frac{N-\delta}{N(N-1)}} \; \sqrt{\frac{\delta}{N}}\right]^T
\end{equation}
and $\bm U\in\mathbb{C}^{N\times N}$ is any unitary matrix with $\bm u$ as its last column. A discussion of the impact of a PAPR constraint, whether solution~\eqref{naif_solution} is adopted or exact solution is numerically determined, is deferred to Sec.~\ref{sec_analysis}

\subsection{Colored radar noise}\label{sec:colored}

The results of the previous sections are based on the assumption that the radar is affected by white noise. In practical systems, the overall interference contains also a fraction of correlated noise, and in fact radar waveform optimization in colored noise is of great interest in radar community~\cite{grossi2012space, tang2010mimo}. Under coherent interference and colored noise case, Problem~\eqref{eq:P0} is reformulated as
%\begin{equation} one-column version
%\label{opt_prob_colored}
%\begin{aligned}
% \max_{\bm s, \bm R_x} & \quad \left\{ \log \det \left( \bm I_N + \frac{|h|^2}{\sigma_v^2} \bm R_x \right) - \beta \log \left( \frac{1 + \frac{\sigma _f^2}{\sigma _v^2} \| \bm s\|^2}{1 + \frac{\sigma _f^2}{\sigma _v^2} \bm s^H \left( \bm I_N + \frac{|h|^2}{\sigma _v^2} \bm R_x \right)^{-1} \bm s} \right) \right\} \\
% \text{s.t.} & \quad \sigma^2_a\bm s^H(\sigma_g^2 \bm R_x + \sigma^2_c\bm s \bm s^H + \bm M)^{-1} \bm s \geq \rho_\text{min}\\
%& \quad \frac{1}{N}\|\bm s\|^2 \leq P_r,  \quad \frac{1}{N}\tr(\bm R_x) \leq P_c, \quad \bm R_x \succeq 0.
%\end{aligned}
%\end{equation}
\begin{equation} % two-column version
\label{opt_prob_colored}
\begin{aligned}
 \max_{\bm s, \bm R_x} & \quad \Bigg\{ \log \det \left( \bm I_N + \frac{|h|^2}{\sigma_v^2} \bm R_x \right)\\
 &\quad \left. - \beta \log \left( \frac{1 + \frac{\sigma _f^2}{\sigma _v^2} \| \bm s\|^2}{1 + \frac{\sigma _f^2}{\sigma _v^2} \bm s^H \left( \bm I_N + \frac{|h|^2}{\sigma _v^2} \bm R_x \right)^{-1} \bm s} \right) \right\} \\
 \text{s.t.} & \quad \sigma^2_a\bm s^H(\sigma_g^2 \bm R_x + \sigma^2_c\bm s \bm s^H + \bm M)^{-1} \bm s \geq \rho_\text{min}\\
& \quad \frac{1}{N}\|\bm s\|^2 \leq P_r,  \quad \frac{1}{N}\tr(\bm R_x) \leq P_c, \quad \bm R_x \succeq 0.
\end{aligned}
\end{equation}
As shown in Appendix~\ref{proof_colored}, this problem admits a solution only if
\begin{equation}
 \rho_\text{min} \leq \frac{\sigma^2_a NP_r}{\sigma^2_c N P_r +\phi_N}\label{existence_colored}
\end{equation}
where $\phi_N$ is the smallest eigenvalue of $\bm M$. In this case, the optimal covariance of the communication system is
\begin{equation}\label{opt_R_colored}
 \bm R_x^* = \bm U^*  \diag \left( \frac{NP_c - \gamma_N^*}{N-1} , \ldots ,  \frac{NP_c - \gamma_N^*}{N-1}, \gamma_N^* \right)  (\bm U^*)^H
\end{equation}
where $\bm U^*\in\mathbb{C}^{N\times N}$ is any unitary matrix whose last column is $\bm v_N$, the latter denoting the eigenvector of $\bm M$ corresponding to $\phi_N$, and $\gamma^*_N$ is reported in Eq.~\eqref{opt_gamma_N_colored} of Appendix~\ref{proof_colored}, while the optimal radar waveform is
\begin{equation}\label{opt_s_colored}
 \bm s^* = \sqrt{ \frac{\rho_\text{min}(\sigma^2_g\gamma_N^*+\phi_N)}{\sigma^2_a-\sigma^2_c\rho_\text{min}}} \bm v_N.
\end{equation}

No closed-form solution is instead available in the case of incoherent interference, and system optimization should rely upon numerical methods.

\subsection{Achievable communication rates}\label{achievable_region}

An important role is played by the region of communication rate pairs, $(R_0, R_1)$, achievable under the constraints of Problem~\eqref{eq:P0}, i.e.,
\begin{align}
{\cal S} (\bm M) & = \bigg\{ \big(R_0(\bm R_x, \bm s), R_1(\bm R_x, \bm s)\big): \notag\\
&\quad \sigma^2_a \bm s^H(\sigma_g^2 \bm R_x + \sigma^2_c \bm s \bm s^H+ \bm M)^{-1} \bm s \geq  \rho_\text{min}, \notag\\
&\quad \frac{1}{N}\tr(\bm R_x) \leq P_c, \frac{1}{N}\|\bm s\|^2 \leq P_r, \bm R_x \succeq 0 \bigg\}.\label{achievable_region_expr}
\end{align}
Knowledge of this region allows determining the optimal transmission policy for any merit function of the form $Q(R_0, R_1)$; since $Q$ is generally increasing in $R_0$ and $R_1$, the solution to the optimization problem would be the point on the border of the achievable region that touches the level set $\left\{ (R_0,R_1): Q(R_0,R_1) = \kappa \right\}$ corresponding to the largest $\kappa \in \mathbb{R}$. Following the proof in Appendix~\ref{proof_lemma_R0_R1_regions}, we have following
\begin{lemma}\label{lemma_R0_R1_regions}
When $\bm R_f = \sigma_f^2 \bm 1_N$, if $(R_0',R_1') \in {\cal S}(\sigma^2_w \bm I_N)$, then there exists a point $(R_0',R_1'') \in {\cal S} (\bm M)$, such that $R_1''\geq R_1'$.
\end{lemma}
According to the lemma, for fixed noise power in radar, white Gaussian noise is the worst case for waveform optimization in the presence of coherent scattering.

\section{Analysis}\label{sec_analysis}

We consider a communication system that, when operating in a non-coexisting mode, may rely on a received Signal-to-Noise Ratio (SNR) per symbol, $|h|^2P_c/\sigma^2_v$, of $10$~dB, so that its maximum rate is simply the capacity, i.e., $\log (1+|h|^2P_c/\sigma^2_v)= 3.46$ bits/channel use, and we set $|h|^2=1$. This is the maximum (un-attainable) rate and is the yardstick we compare our results to. As to the radar system, we assume it is designed so that, when operating in non-coexisting mode, under white noise and no clutter, it detects a reference target, located at the $K$-th range cell, and whose RCS is exponentially distributed with average value $\sigma^2_a=1$~m$^2$, with probability of false alarm ($P_\text{fa}$) and probability of detection ($P_\text{d}$) equal to $10^{-4}$ and $0.9$, respectively. Since, for such a Swerling~I target, $P_\text{d}=P_\text{fa}^{1/(1+NP_r\sigma^2_a/\sigma^2_w)}$~\cite{Richards_2005}, this conditions corresponds to requiring a \emph{cumulated} SNR $NP_r\sigma^2_a/\sigma^2_w=19.4$~dB. Similar to $|h|^2$, we assume $\sigma^2_g$, the variance of the coupling coefficient from the communication transmitter to the radar receiver, equal to $1$. Concerning the interference, we define, at the communication system, the Interference-to-Noise Ratio (INR) as $\sigma^2_f/\sigma^2_v$ and, at the radar, the Signal-to-Clutter Ratio (SCR) as $\sigma^2_a/\sigma^2_c$.

\begin{figure}[t]
 \centering
% \centerline{\includegraphics[width=0.6\columnwidth]{fig_04.pdf}} % one-column version
\centerline{\includegraphics[width=\columnwidth]{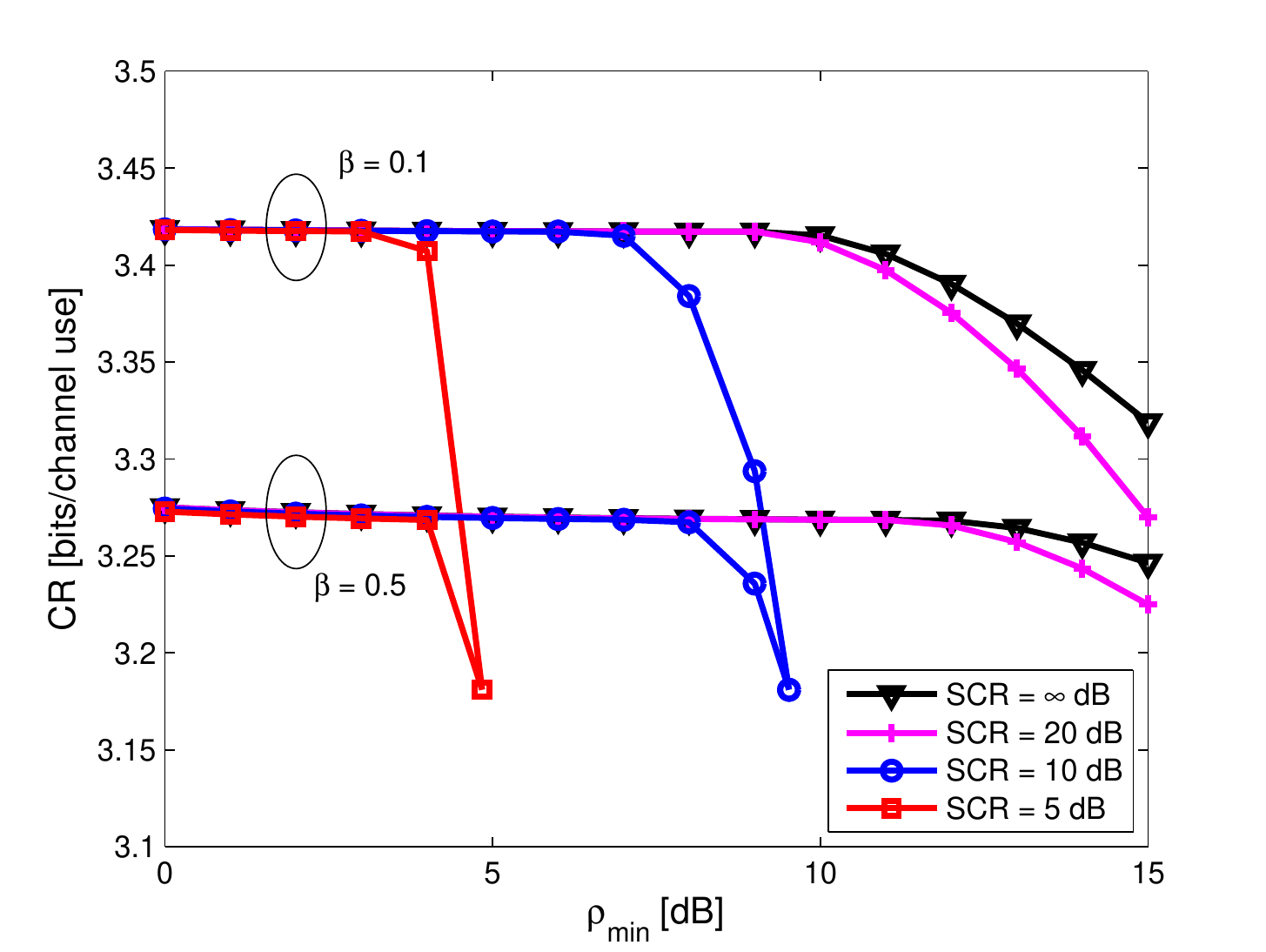}} % two-column version
 \caption{Optimized compound rate versus the minimum required SINR for two values of $\beta$ when $N=8$, $\text{INR}=10$~dB, $\alpha=\beta$, and the radar noise is white.} \label{fig_04}
\end{figure}

At first, we study the impact, on the communication system, of the constraint forced on the minimum SINR received by the radar. The results, referring to the cases $\beta=0.1$ and $\beta=0.5$ (representing scattering environments with different densities) are represented in Fig.~\ref{fig_04}, where the optimum CR is plotted versus $\rho_\text{min}$ for different SCR's when $N=8$, $\text{INR}=10$~dB, $\alpha=\beta$, and the radar noise is white. We underline here that, in the considered scenario, the limiting performance of the jointly optimal design under coherent and incoherent scattering is the same, whereby the curves corresponding to the optimum CR is unique. Not surprisingly, larger values of $\beta$ turn out to be detrimental in terms of CR; the effect of SCR, instead, is rather dramatic, because it limits the feasibility region of the optimum design, mainly because the SINR constraint cannot be met.

\begin{figure}[t]
 \centering
% \centerline{\includegraphics[width=0.6\columnwidth]{fig_05.pdf}} % one-column version
\centerline{\includegraphics[width=\columnwidth]{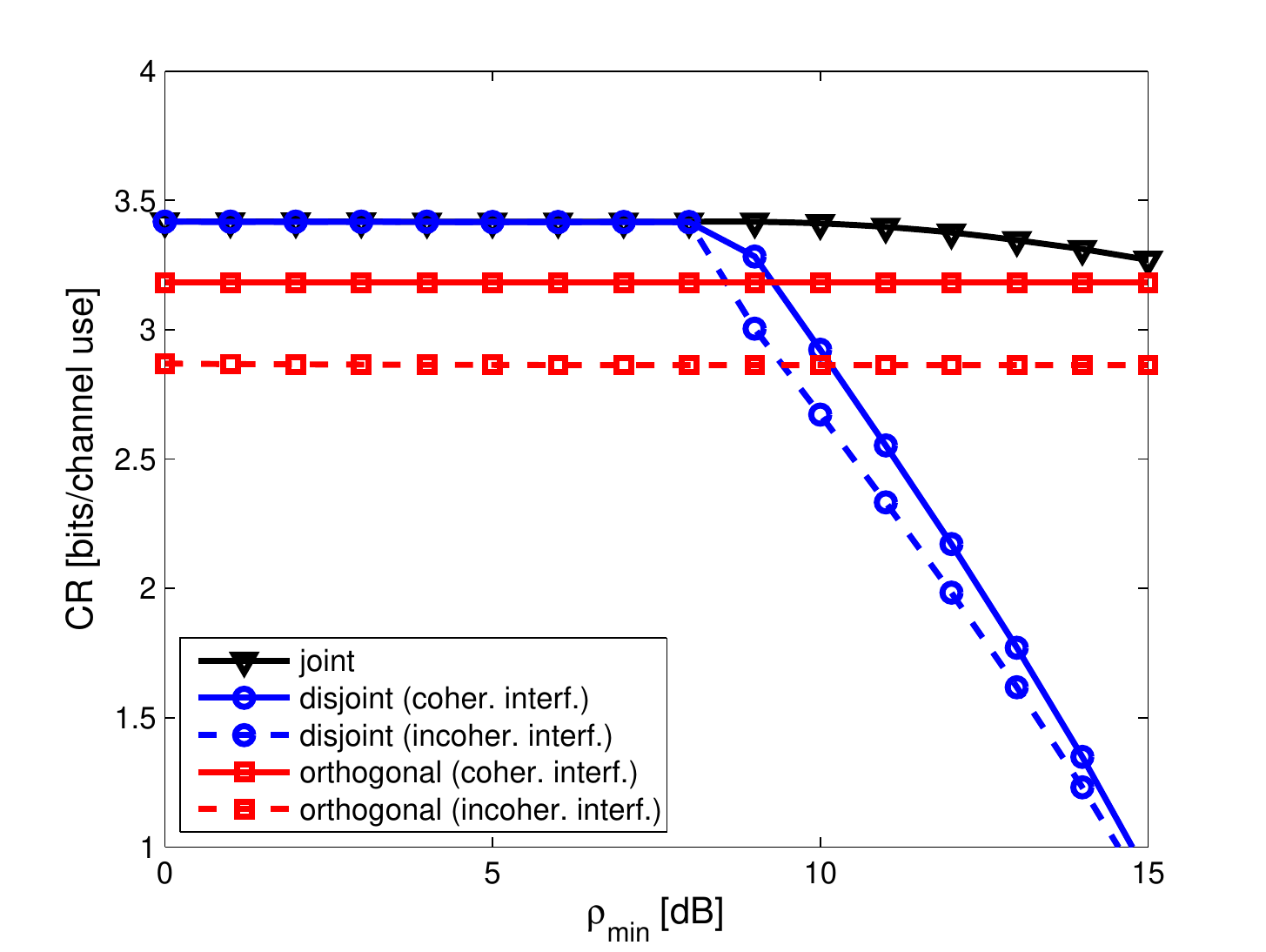}} % two-column version
 \caption{Compound rate versus the minimum required SINR for the optimum design in~\eqref{eq:P0}, the orthogonal design (where systems transmit in orthogonal subspaces), and the disjoint design (where the systems ignore coexistence), when $N=8$, $\text{INR}=10$~dB, $\text{SCR}=20$~dB, $\beta=\alpha=0.1$, and the radar noise is white; both coherent and incoherent interference are considered.} \label{fig_05}
\end{figure}

The advantages of a joint optimized design over a non-cooperative approach are outlined in Fig.~\ref{fig_05}; here we have reported the optimal joint design, the disjoint design, where both systems optimize the respective performance measure ignoring coexistence,\footnote{In this case, the communication system uses the correlation matrix $\bm R_x =P_c \bm I_N$, while the radar, after having estimated the overall disturbance, employes an unmodulated pulse train with the minimum power satisfying the SINR constraint.} and the orthogonal design, where the two systems transmit in orthogonal spaces; the other parameters are $N=8$, $\text{INR}=10$~dB, $\text{SCR}=20$~dB, $\alpha=\beta=0.1$, and white radar noise. The results clearly indicate the disjoint design is nearly optimal for small values of $\rho_\text{min}$, while being catastrophic at larger values of $\rho_\text{min}$.  Also, while for coherent scattering, the conservative approach of transmitting into orthogonal subspaces guarantees a performance level very close to that of the joint design, this is no longer the case for incoherent scattering, since the reflector scintillations spreads the interference at the communication systems over all of the directions of the signal space. Similar trends are observed for higher values of $\beta=\alpha$, even though we do not show this curves for space limitation.

\begin{figure}[t]
\centering
% \centerline{\includegraphics[width=0.6\columnwidth]{fig_06.pdf}} % one-column version
\centerline{\includegraphics[width=\columnwidth]{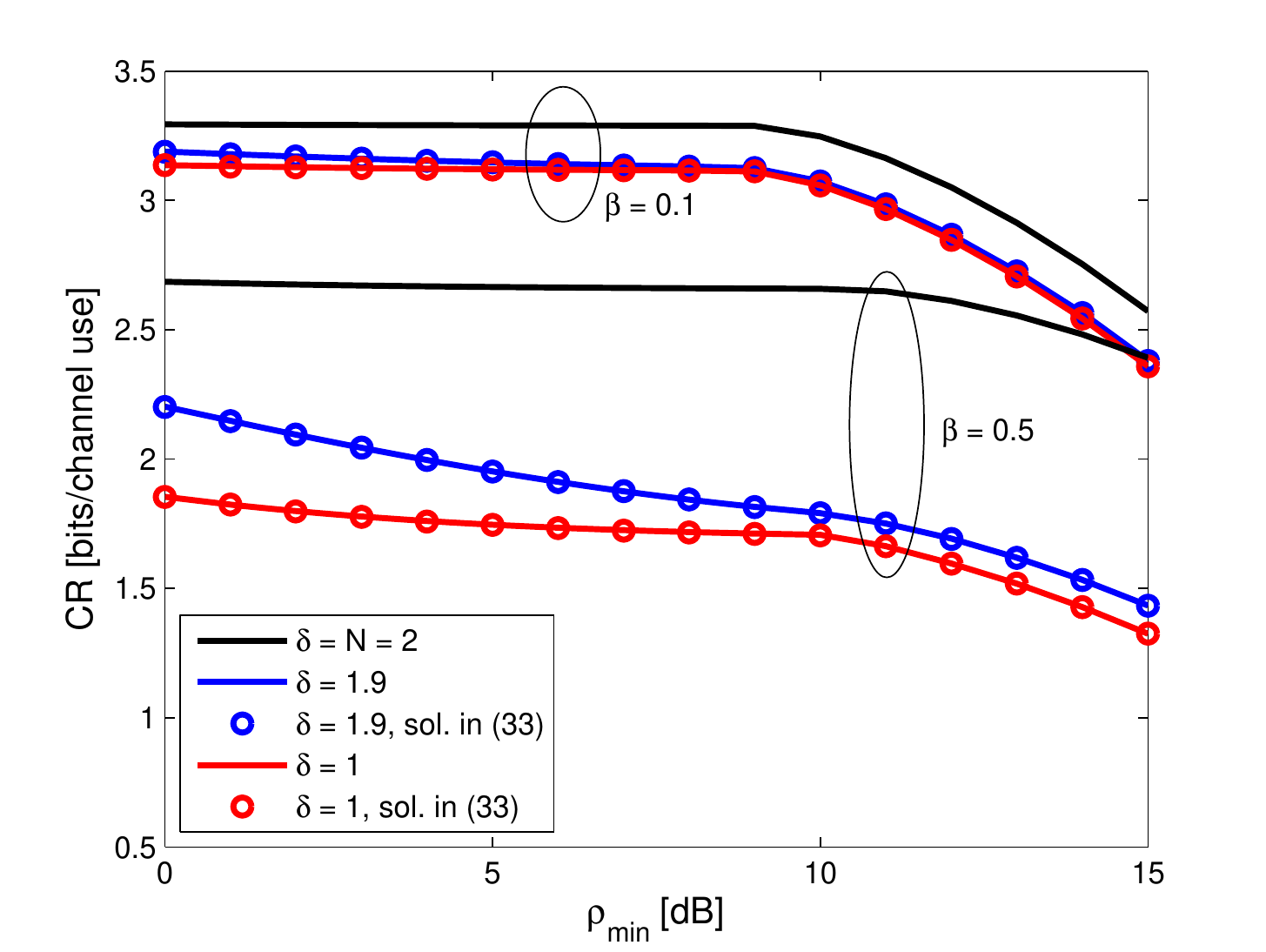}} % two-column version
 \caption{Optimized compound rate versus the minimum required SINR for incoherent interference in the presence of a PAPR constraint $\delta$ for two values of $\beta$ when $N=2$, $\text{INR}=10$~dB, $\text{SCR}=20$~dB, $\alpha=\beta$, and the radar noise is white. The case $\delta=N$ corresponds to unconstrained PAPR. For comparison purposes, the na\"if solution in~\eqref{naif_solution} is also included.} \label{fig_06}
\end{figure}

\begin{figure}[t]
 \centering
% \centerline{\includegraphics[width=0.6\columnwidth]{fig_07.pdf}} % one-column version
\centerline{\includegraphics[width=\columnwidth]{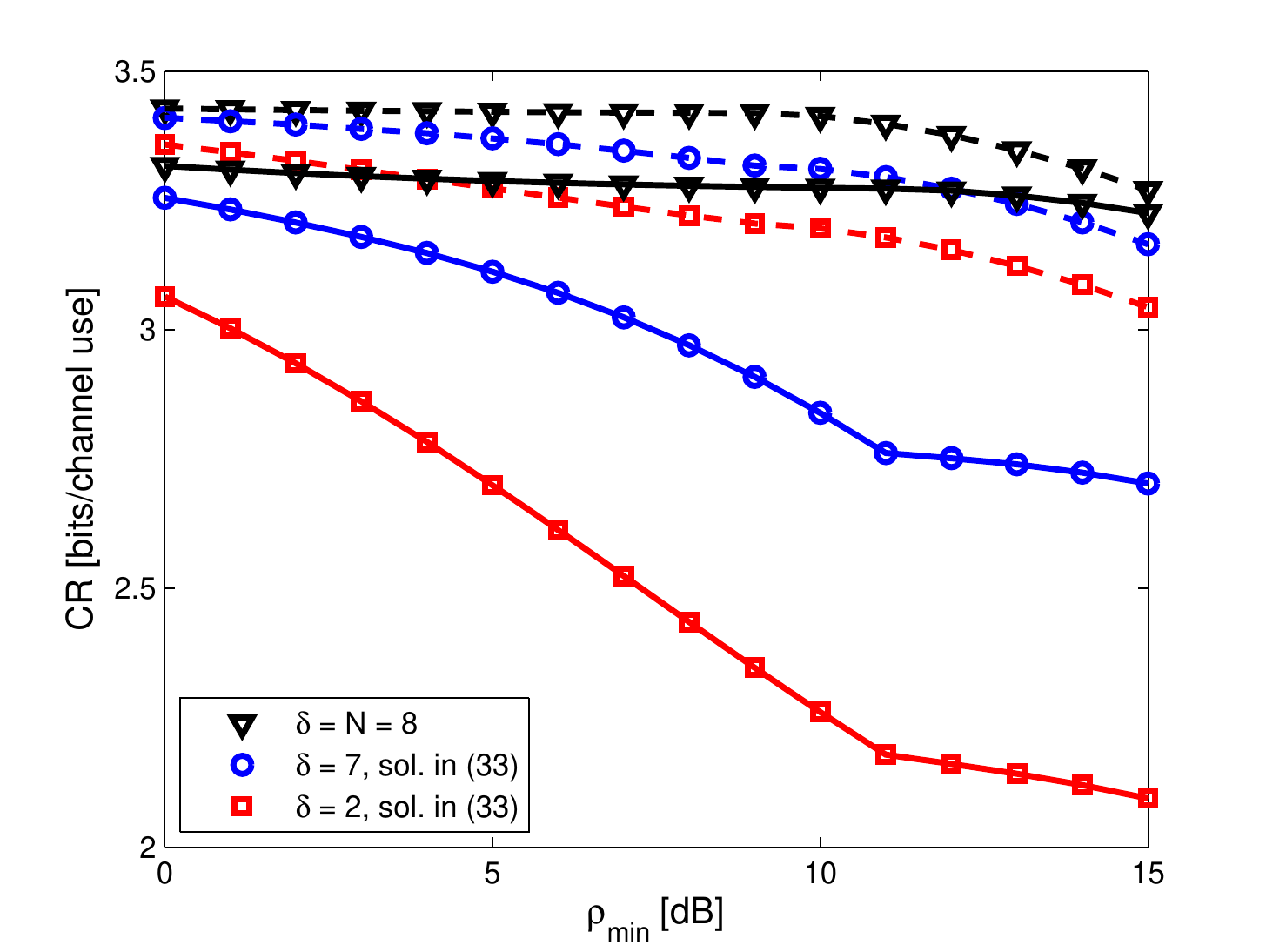}} % two-column version
 \caption{Compound rate versus the minimum required SINR for incoherent interference in the presence of a PAPR constraint $\delta$ when $N=8$, $\text{INR}=10$~dB, $\text{SCR}=20$~dB, $\alpha=\beta$, and the radar noise is white. Dashed lines refer to $\beta = 0.1$ and solid lines to $\beta =0.5$; the case $\delta=N$ corresponds to unconstrained PAPR.} \label{fig_07}
\end{figure}

Next, we analyze the effect of an additional PAPR constraint: as we have seen in Sec.~\ref{sec:codesign}, this does not affect the design in the presence of coherent interference, where the eigenvector matrix can be freely chosen, but it may influence the design in the presence of incoherent interference. In the latter case, we resort to numerical methods to derive the solution, and we compare it with the ``na\"if'' solution in~\eqref{naif_solution}. The results are shown in Figs.~\ref{fig_06}, where the optimized CR is reported versus $\rho_\text{min}$ for two values of $\delta$ and $\beta$, when $\text{INR}=10$~dB, $\text{SCR}=20$~dB, and $\alpha=\beta$, assuming white radar noise. In order to reduce the computational burden, we have set $N=2$. It can be seen that the solution in~\eqref{naif_solution} results in a CR almost coincident with the optimal one, even for the tightest PAPR constraint, $\delta=1$, corresponding to a constant amplitude pulse train. We therefore use the solution in~\eqref{naif_solution} to study the trends in CR performance at higher values of $N$. In Fig.~\ref{fig_07} the optimized CR is reported versus $\rho_\text{min}$ for $\beta=0.1$ and $\beta=0.5$, when $N=8$, $\text{INR}=10$~dB, $\text{SCR}=20$~dB, and $\alpha=\beta$, assuming white radar noise. Obviously, more stringent constraints on the PAPR (i.e., smaller values of $\delta$) result in larger losses in terms of CR with respect to the case where PAPR is unconstrained ($\delta=N=8$). However, the sensitivity of the performance to the PAPR constraint is modest as far as $\delta$ is large enough to allow a substantial reduction of the amount of the interference along the $N-1$ dimensions that the optimum solution would guarantee interference-free.

\begin{figure}[t]
 \centering
%  \centerline{\includegraphics[width=0.6\columnwidth]{fig_08.pdf}} % one-column version
 \centerline{\includegraphics[width=\columnwidth]{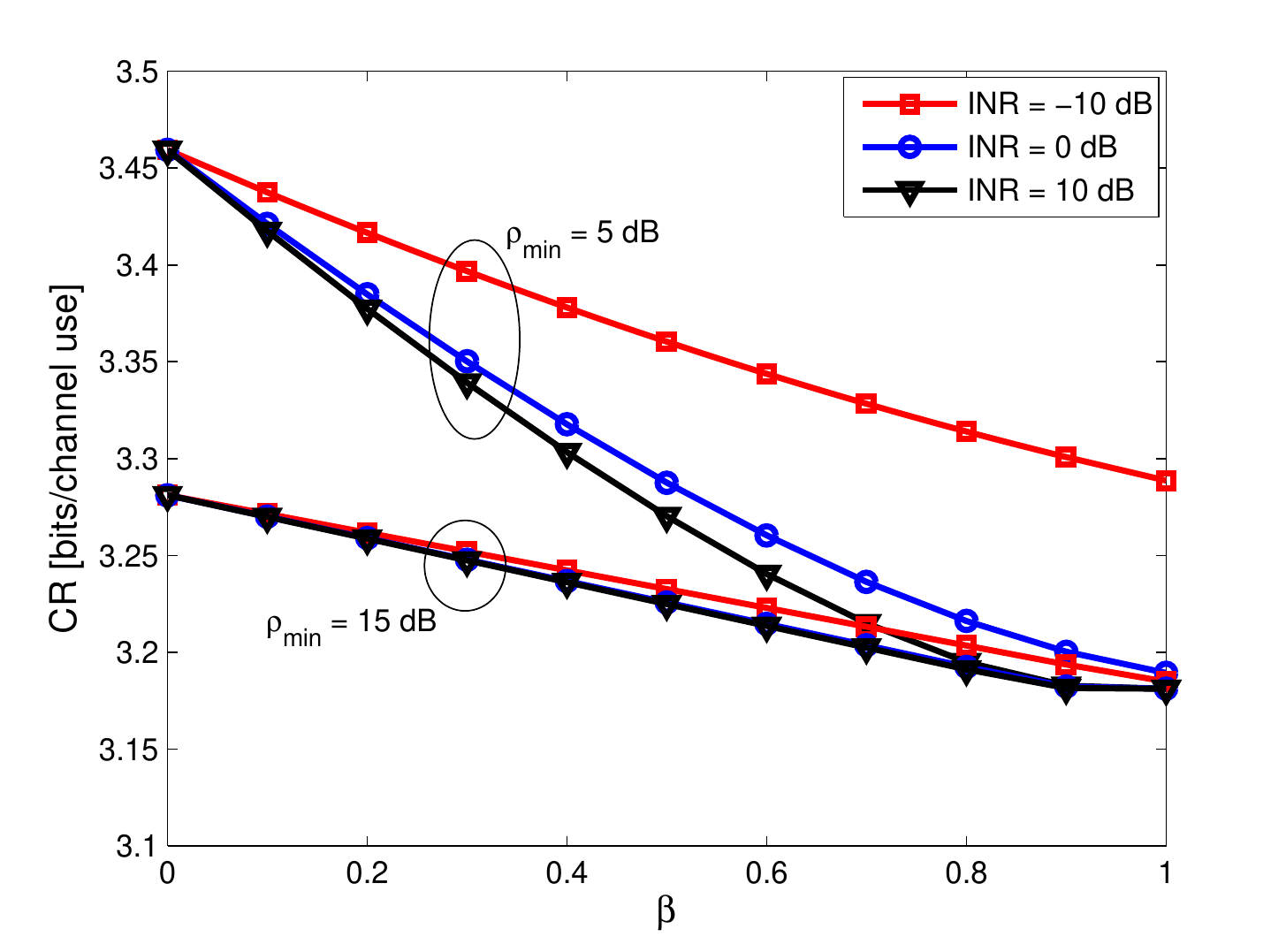}} % two-column version
  \caption{Optimized CR versus $\beta$ for different values of INR and $\rho_\text{min}$, when $N=8$, $\text{SCR}=20$~dB, $\alpha=\beta$, and the radar noise is white.} \label{fig_08}
\end{figure}

\begin{figure}[t]
 \centering
%  \centerline{\includegraphics[width=0.6\columnwidth]{fig_09.pdf}} % one-column version
 \centerline{\includegraphics[width=\columnwidth]{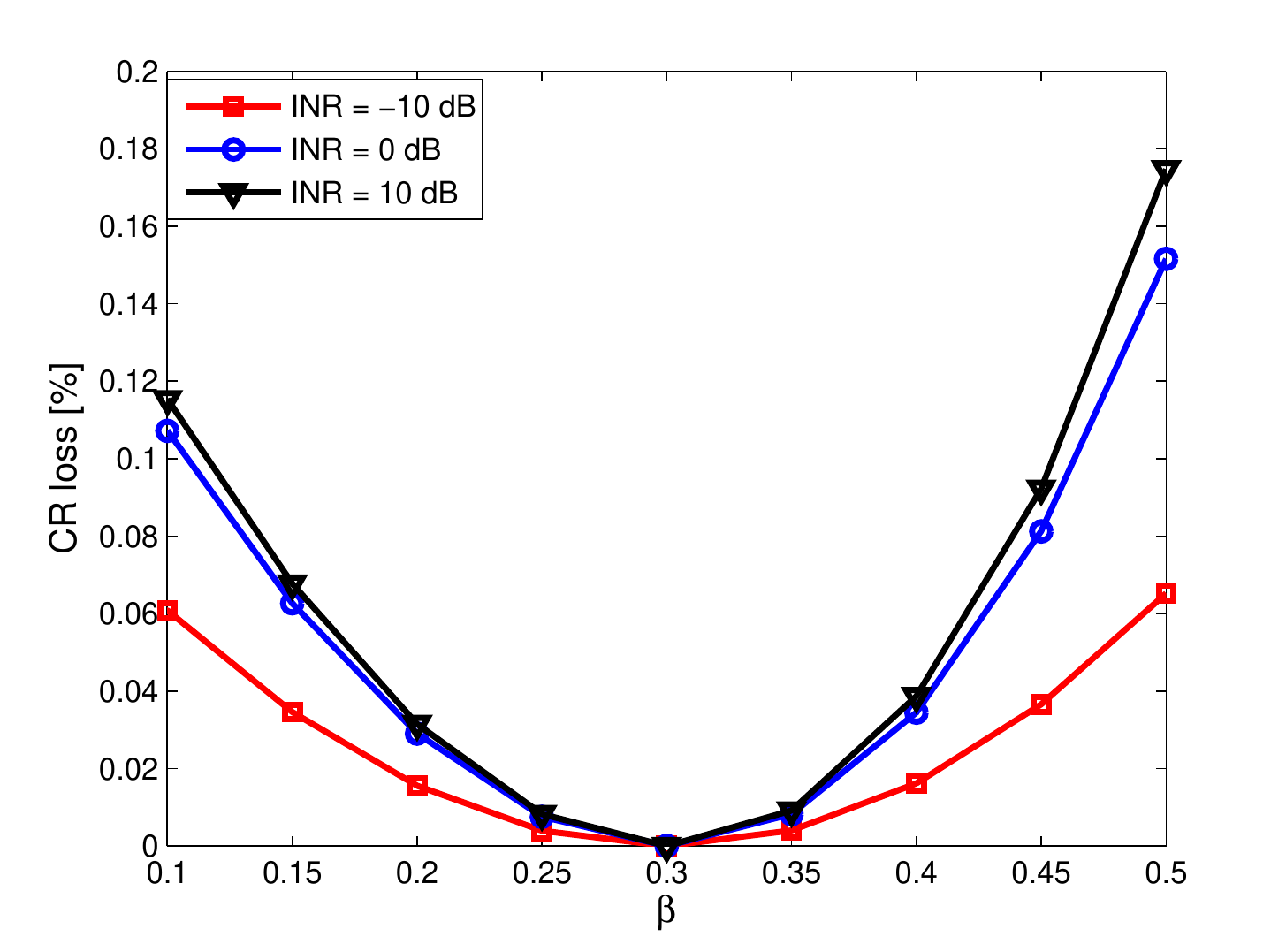}} % one-column version
  \caption{Compound rate loss versus $\beta$ for $\rho_\text{min}=5$~dB when $\alpha=0.3$, $N=8$, $\text{INR}=10$~dB, and $\text{SCR}=20$~dB.} \label{fig_09}
\end{figure}

The effect of $\beta$ is highlighted in Fig.~\ref{fig_08} for different values of INR and two values of $\rho_\text{min}$, when $N=8$, $\text{SCR}=20$~dB, $\alpha=\beta$, and the radar noise is white. As expected, increasing values of $\beta$ result in decreasing values of the optimized CR, on the understanding that the loss is at most in the order of $1/N$. The effects of possible mismatches between $\alpha$ (the true interference density) and $\beta$ (the value assumed at the design stage) is elicited in Fig.~\ref{fig_09}, referring to the cases of $\rho_\text{min}=5$~dB, $\alpha=0.3$, $N=8$, $\text{INR}=10$~dB, and $\text{SCR}=20$~dB. As it can be seen, the percentage loss is very limited, which shows marked robustness of the proposed joint design scheme with respect to possible errors in the estimated target density. The percentage loss when $\rho_\text{min}=10$~dB is slightly smaller, the figure not being reported due to lack of space.

\begin{figure}[t]
 \centering
% \centerline{\includegraphics[width=0.6\columnwidth]{fig_10.pdf}} % one-column version
\centerline{\includegraphics[width=\columnwidth]{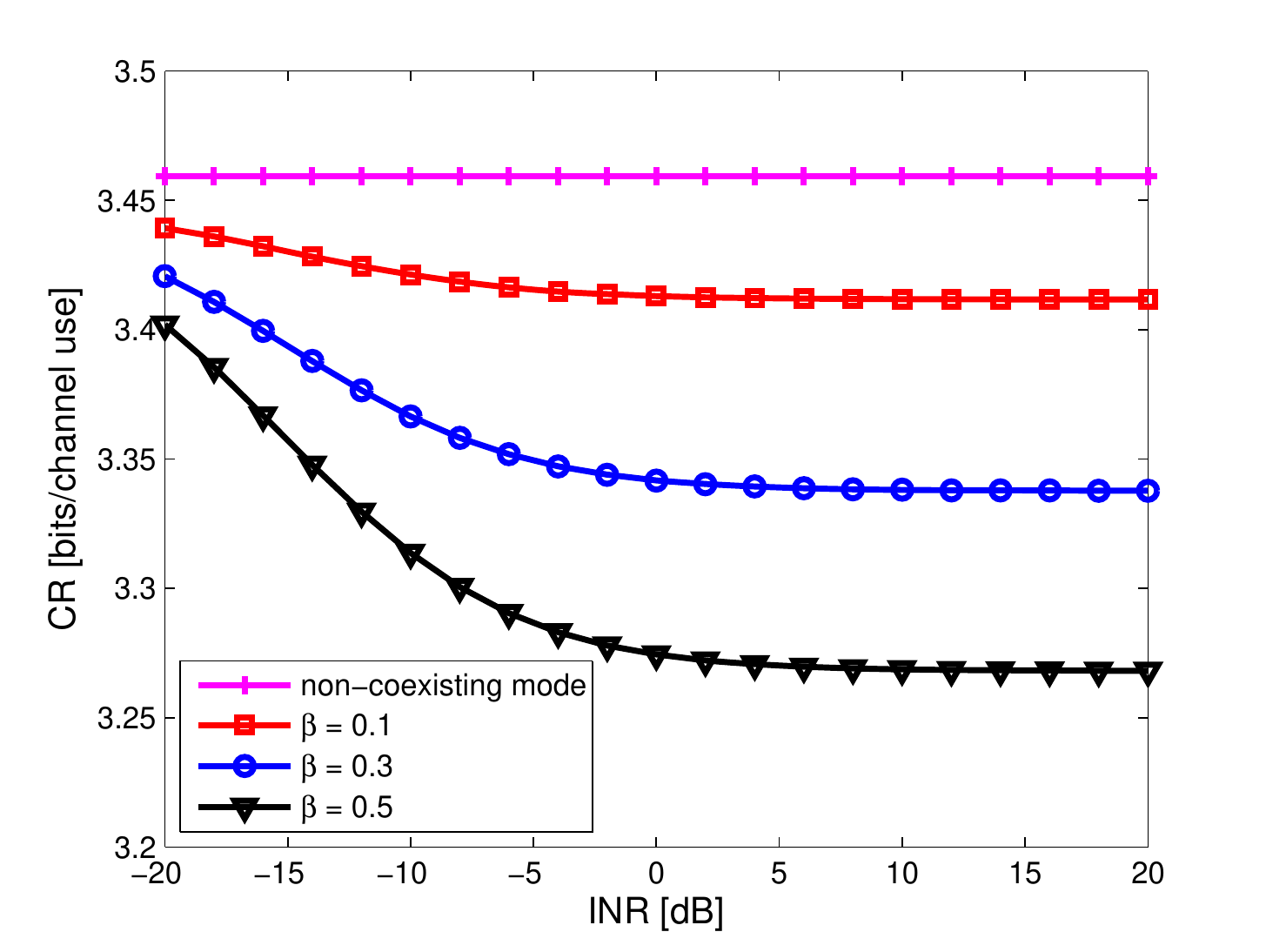}} % two-column version
  \caption{Optimized compound rate versus the interference-to-noise ratio with different values of $\beta$ when $\rho_\text{min}=10$~dB, $N=8$, $\text{SCR}=20$~dB, and $\beta=\alpha$; for comparison purposes, the non-coexisting case is also included.} \label{fig_10}
\end{figure}

Fig.~\ref{fig_10} represents the optimized CR for different values of $\beta$ and varying INR. It is interesting to observe that the ``strong interference'' condition is reached pretty soon, i.e., for INR in the order of 0~dB.

\begin{table}[t]
 \caption{Optimized CR [bits/channel use] for $\rho_\text{min}=10$~{\upshape dB}, $\text{SCR}=20$~{\upshape dB}, $\beta=\alpha$, and white radar noise}
 \label{table_CR}
 \centering
 \begin{tabular}{ccccc}
 \toprule
  &\multicolumn{2}{c}{$\text{INR} = -10$~dB}& \multicolumn{2}{c}{$\text{INR} = 10$~dB} \\
  \cline{2-5}
  $N$ & $\beta=0.1$ & $\beta = 0.5$ &  $\beta=0.1$ &  $\beta = 0.5$\\
  \midrule
  2 & 3.29 & 2.85 & 3.25 & 2.66 \\
  4 & 3.38 & 3.16 & 3.36 & 3.07 \\
  8  & 3.42 & 3.31 & 3.41 & 3.27 \\
  16 & 3.44 & 3.39 & 3.44 & 3.36 \\
  32 & 3.45 & 3.42 & 3.45 & 3.41 \\
\bottomrule
 \end{tabular} \\
\end{table}

The effect of the pulse number onto the optimized CR is instead studied in Table~\ref{table_CR} when $\rho_\text{min}=10$~dB, $\text{SCR}=20$~dB, $\beta=\alpha$, and the radar noise is white. This table should be read and interpreted in the light of~\eqref{eq:bound_mutual}. Indeed, while increasing with $N$, CR is not a rate achievable through the proposed encoding scheme: however, since its distance with respect to the MI per channel use decreases as $\frac{1}{N}$, Table~\ref{table_CR} gives an idea of the transmission rates achievable by the communication system through Gaussian random coding employing the matrix $\bm R_x$ for varying $\beta$ and interferers strength. For example, we see that, if 10\% of the scatterers interfer with the communication system, then using a pulse train of length $32$ makes it possible to achieve a transmission rate in the range $[3.34, 3.45]$ bits/channel use\footnote{We remind here however that $N=32$ could not be enough to enforce the asymptotic behavior the channel coding theorem relies upon. All the subsequent considerations should be read in the light of this fact.} under both weak and strong interferers; this interval becomes $[2.86, 3.27]$ bits/channel use for $N=8$, $\beta=0.5$ and strong interference.

\begin{figure}
 \centering
% \centerline{\includegraphics[width=0.6\columnwidth]{fig_11.pdf}} % one-column version
\centerline{\includegraphics[width=\columnwidth]{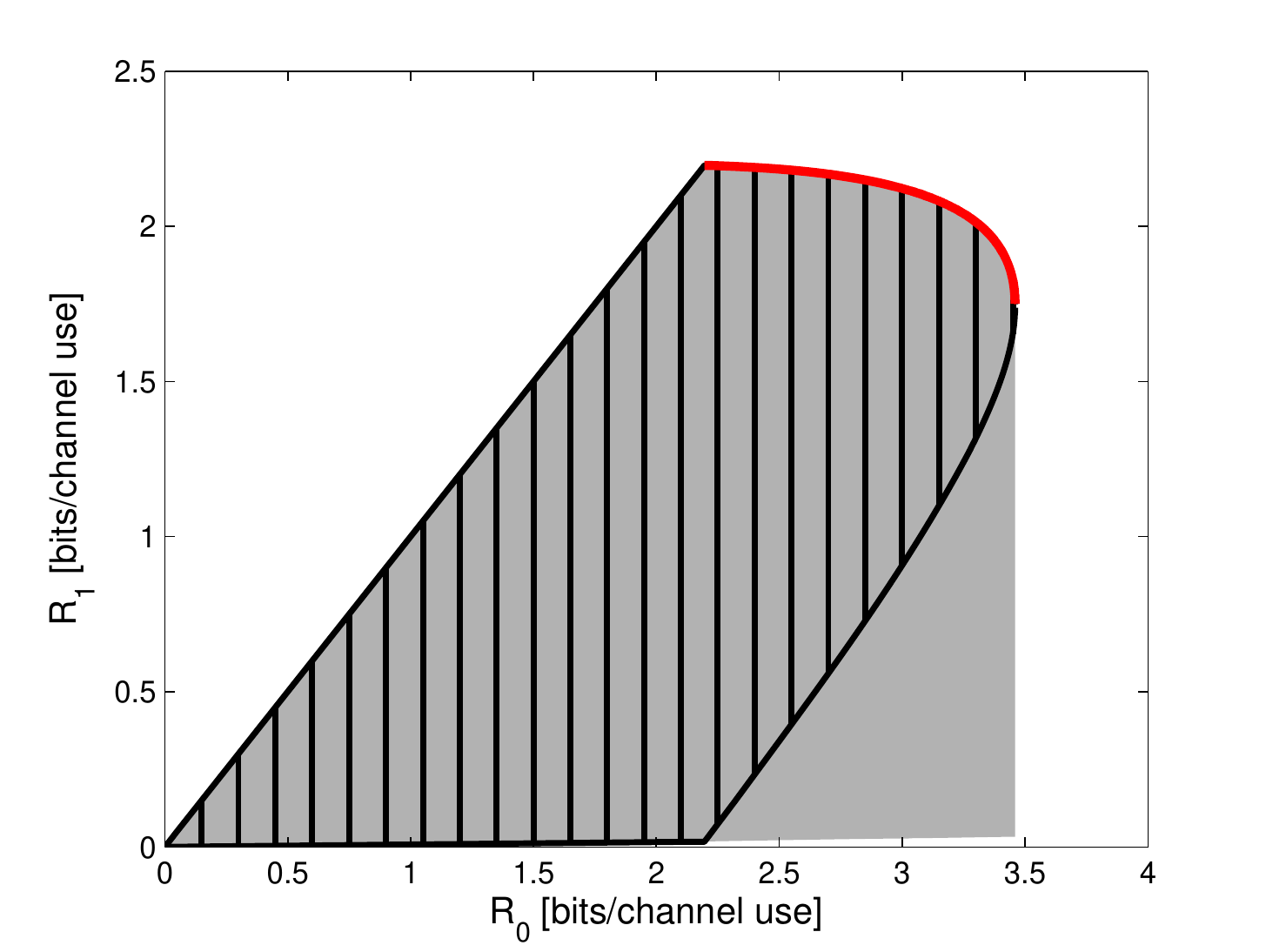}} % two-column version
 \caption{Example of the achievable region of rate pairs ${\cal S}$ in~\eqref{achievable_region_expr} (grey area corresponds to incoherent interference, hatched area to coherent interference) and of the curve $\psi$ in~\eqref{function_psi} (red line), when $\rho_\text{min}=5$~dB, $\text{INR}=10$~dB, $\text{SCR}=20$~dB, $\beta=\alpha$, and the radar noise is white.} \label{fig_11}
\end{figure}

Finally, we focus our attention on the region of achievable rate pairs, ${\cal S}(\bm M)$, defined in~\eqref{achievable_region_expr}, which, as remarked in Sec.~\ref{achievable_region}, is key for determining the transmission policy optimizing arbitrary merit functions of the form $Q(R_0,R_1)$: in particular, the maximum of any \emph{reasonable} $Q(R_0,R_1)$ will lie on the upper boundary of ${\cal S}(\bm M)$. Therefore, we also introduce the curve $\psi(\,\cdot\,,\bm M):[0,1]\rightarrow \mathbb{R}^2$, defined as
\begin{equation} \label{function_psi}
 \psi(\beta, \bm M) = \argmax_{(R_0, R_1) \in {\cal S}(\bm M)} \big\{\beta R_1 + (1-\beta) R_0\big\}
\end{equation}
which lies on such boundary. In Fig.~\ref{fig_11}, the region ${\cal S}$ and the curve $\psi$ are reported for coherent and incoherent interference when $\rho_\text{min}=5$~dB, $\text{INR}=10$~dB, $\text{SCR}=20$~dB, $\beta=\alpha$, and the radar noise is white. In order to  reduce the computational complexity of the simulations, we have set $N=2$. It can be seen that the region corresponding to coherent interference is included into the one corresponding to incoherent interference, and that the upper border is coincident in the two cases, which also confirms the result of Sec.~\ref{sec:incoher_design}, stating that the optimized CR's are equal. 

\begin{figure}
 \centering
%  \centerline{\includegraphics[width=0.6\columnwidth]{fig_12.pdf}} % one-column version
 \centerline{\includegraphics[width=\columnwidth]{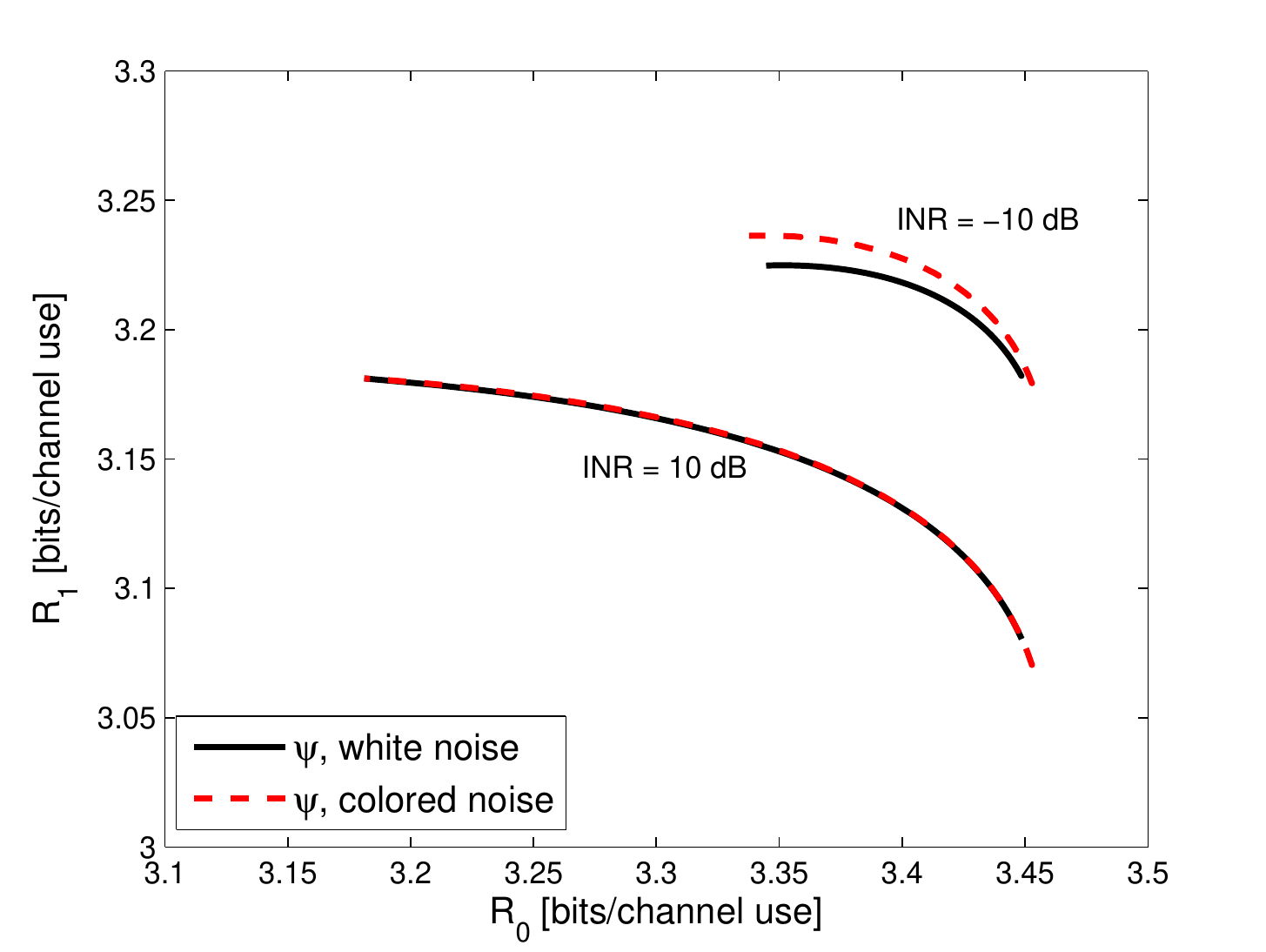}} % two-column version
 \caption{Plot the curve $\psi(\,\cdot\,, \bm M)$ in~\eqref{function_psi} for two values of the INR when $\bm M=\sigma^2_w \bm I_N$ (white radar noise) and when $M_{i,j}=\sigma^2_w (2/3)^{|i-j|}$(exponentially correlated radar noise); the remaining parameters are $N=8$, $\rho_\text{min}=10$~dB, $\text{SCR}=20$~dB, and $\beta=\alpha$.} \label{fig_12}
\end{figure}

The impact of the correlation of the interference impinging on the radar is finally studied in Fig.~\ref{fig_12}, where the curves $\psi(\,\cdot\,, \bm M)$ when $\bm M=\sigma^2_w \bm I_N$ (white radar noise) and when $\bm M$ is such that $M_{i,j}=\sigma^2_w 2^{-|i-j|}$ (exponentially correlated radar noise) is reported; the remaining parameters are $N=8$, $\rho_\text{min}=10$~dB, $\text{SCR}=20$~dB, $\beta=\alpha$, and two values of INR are considered. As proven in Lemma~\ref{lemma_R0_R1_regions}, white radar noise turns out to be the most detrimental situation, even though this is relevant for low values of INR.

\section{Conclusions}\label{sec_conclusion}

This paper has introduced a new framework for co-existing radar and communication systems. While the key performance measure for the radar remains the SINR, the performance of the communication system is characterized through a new figure of merit, the Compound Rate, which reflects the intermittent nature of the interference generated by the radar and, form an information-theoretic point of view, is a conditional mutual information of a channel which may be or not subject to interference. This is a direct consequence of the fact that a communication system co-existing with a pulsed radar behaves as a set of independent parallel channels affected by contaminated-normal interference. Considering arbitrary correlation of the radar interference and two fundamental situations of interference generated by the radar (i.e., perfectly coherent and totally incoherent interference), we illustrate the design of the radar waveform and the communication system encoding matrix by maximizing the compound rate under a constraint on the SINR of the radar, showing that co-design is key in order to guarantee the performance of both systems in co-existing architectures.

The proposed setup, which has been developed, at the radar side, assuming slow-time coding only, lends itself to incorporate also fast-time coding. Also, the scheme assumes frame synchronism between the radar and the communication receiver: even though such an assumption is rather mild in the considered scenario, current efforts of the authors are directed towards the design of timing-free schemes, which would allow extending the present framework to multiple co-existing systems, e.g., involving a number of radars, using different waveforms and occupying different locations, co-existing with one (or more) communication systems. Finally, the impact of some cognition of the surrounding scene on the achievable performance could be a topic of great interest, since it would allow releasing the ``worst case'' philosophy in favor of further optimization.

\section*{Acknowledgment}
The Authors would like to thank Augusto Aubry, University of Naples ``Federico~II,'' for constructive comments and discussions.

\appendix

This Appendix contains six sections. In next section we report the connection between the SINR in~\eqref{SINR} and the Kullback-Leibler divergence pair the observations in~\eqref{eq:radar-1}. In Secs.~\ref{proof_coherent},~\ref{proof_incoherent}, and~\ref{proof_colored}, we derive the solutions to the optimization Problems in~\eqref{opt_prob_coher},~\eqref{opt_prob_incoher}, and~\eqref{opt_prob_colored}, respectively. Finally, in Sec.~\ref{proof_lemma_R0_R1_regions}, we report the proof of Lemma~\ref{lemma_R0_R1_regions}. Before proceeding, we give the following simple lemma, which will be used next.
\begin{lemma}\label{SINR_bound}
 Let $\bm y \in\mathbb{C}^N$ and $\bm A \in \mathbb{C}^{N\times N}$ be an Hermitian positive definite matrix; let $\lambda_\text{min}(\,\cdot\,)$ and $\bm u_\text{min}(\, \cdot\,)$ denote the smallest eigenvalue and the corresponding eigenvector, respectively, of the matrix in the parentheses; then we have
\begin{equation}
 \bm y^H  (\bm A  + \bm y \bm y^H )^{-1} \bm y \leq \frac{ \| \bm y \|^2}{\lambda_\text{min}(\bm A) + \| \bm y \|^2}
\end{equation}
and equality holds if $\bm u_\text{min}(\bm A)$ and $\bm y$ are linearly dependent.
\end{lemma}
\begin{IEEEproof}
Exploiting~\cite{Mirsky1975}, we have
\begin{align}
 \bm y^H \left( \bm A + \bm y \bm y^H  \right)^{-1} \bm y & = \tr\left(\left( \bm A + \bm y \bm y^H  \right)^{-1} \bm y \bm y^H \right)\notag\\
 & \leq \frac{\| \bm y \|^2}{\lambda_\text{min}\left( \bm A  + \bm y \bm y^H \right)}\notag\\
 & \leq \frac{ \| \bm y \|^2}{\lambda_\text{min}(\bm A) + \| \bm y \|^2}
\end{align}
where the last inequality follows from the property of the minimum operator, and equality holds if $\bm u_\text{min}(\bm A)$ and $\bm y$ are linearly dependent.
\end{IEEEproof}

\subsection{Kullback-Leibler divergence and SINR}\label{proof_KL_div}

Denoting with $f_1$ and $f_0$ the densities of the observations in~\eqref{eq:radar-1} under $H_1$ and $H_0$, respectively, we have that, when $g$ is deterministic and $a\sim {\cal CN}(0,\sigma^2_a)$,
\begin{subequations}
\begin{align}
f_0 \left( \bm r \right) &= \frac{e^{-\tr\left( \bm r^H (\sigma_g^2 \bm R_x  + \sigma^2_c \bm s \bm s^H + \bm M)^{-1} \bm r \right)}}{ \pi^N \det(\sigma_g^2 \bm R_x  + \sigma^2_c \bm s \bm s^H + \bm M)} \\
f_1 \left( \bm r \right) &= \frac{e^{-\tr\left( \bm r ^H (\sigma_g^2 \bm R_x  + \sigma^2_c \bm s \bm s^H + \bm M + \sigma_a^2 \bm s \bm s^H )^{-1} \bm r \right)}}{ \pi^N \det(\sigma_g^2 \bm R_x  + \sigma^2_c \bm s \bm s^H + \bm M + \sigma_a^2 \bm s \bm s^H)}
\end{align}%
\end{subequations}
so that the log-likelihood ratio of $f_1$ to $f_0$ is
\begin{align}
\log \frac{f_1 (\bm r)}{f_0 (\bm r)} &= \tr \Big( \bm r ^H \big( (\sigma_g^2 \bm R_x +  \bm M)^{-1} \notag\\
&\quad  - (\sigma_g^2 \bm R_x +  \bm M + \sigma_a^2 \bm s \bm s^H)^{-1} \big) \bm r \Big) \notag \\
&\quad - \log \det \Big( (\sigma_g^2 \bm R_x  + \sigma^2_c \bm s \bm s^H + \bm M + \sigma_a^2 \bm s \bm s^H) \notag\\
&\quad \times (\sigma_g^2 \bm R_x  + \sigma^2_c \bm s \bm s^H + \bm M)^{-1} \Big).
\end{align}
Then, it can be easily shown that the Kullback-Leibler divergences between these two densities are
\begin{subequations}
 \begin{align}
 D(f_1 \Vert f_0) &= \SINR - \log (1 + \SINR)\\
 D(f_0 \Vert f_1) &= \log (1 + \SINR) -\frac{\SINR}{1+\SINR}
 \end{align}%
\end{subequations}
which are both increasing with $\SINR$. Therefore, maximizing the SINR is equivalent to maximizing the divergences.

\subsection{Solution to Problem~\eqref{opt_prob_coher}}\label{proof_coherent}

We present the proofs for the three situations in Sec.~\ref{sec:coher_design}.

\subsubsection{Fixed communication codebook}\label{proof_coher_fixed_R}

Let $\lambda_N$ and $\bm u_N$ be the smallest eigenvalue of $\bm R_x$ and the corresponding eigenvector, respectively. Then, from Lemma~\ref{SINR_bound}, we have
\begin{equation}
 \sigma^2_a\bm s^H(\sigma_g^2 \bm R_x + \sigma^2_c\bm s \bm s^H + \sigma^2_w \bm I_N)^{-1} \bm s \leq \frac{\sigma^2_a \| \bm s \|^2}{\sigma_g^2 \gamma_N + \sigma^2_c \| \bm s\|^2 + \sigma^2_w}
\end{equation}
and equality holds if $\bm s = \|\bm s\| \bm u_N$. Therefore, the problem admits a solution only if condition~\eqref{existence_coher_fixed_R} is satisfied. In this case, for fixed $\| \bm s \|^2= \epsilon$, the solution is $\bm s = \sqrt{\epsilon} \bm u_N$, for it guarantees the largest SINR level and, again from Lemma~\ref{SINR_bound}, the largest CR, for the latter is increasing with $\bm s^H\big(\bm I_N +\frac{|h|^2}{\sigma^2_v} \bm R_x \big)^{-1} \bm s$. As to $\epsilon$, it must be determined by solving
\begin{equation}
\begin{aligned}
\min_{ \epsilon} & \quad \frac{1+\frac{\sigma _f^2}{\sigma^2_v} \epsilon}{1+ \left(1  + \frac{|h|^2}{\sigma^2_v} \gamma_N \right)^{-1} \frac{\sigma _f^2}{\sigma^2_v} \epsilon}\\
\text{s.t.} & \quad  \frac{\rho_\text{min}(\sigma^2_g\gamma_N+\sigma^2_w)}{\sigma^2_a-\sigma^2_c\rho_\text{min}} \leq \epsilon \leq NP_r.
\end{aligned}
\end{equation}
Since the objective function is increasing, the solution is the lower bound for $\epsilon$, and the optimal $\bm s$ is the one reported in~\eqref{opt_s_coher_fixed_R}.

\subsubsection{Fixed radar waveform}\label{proof_coher_fixed_s}

Suppose that $\bm R_x$ satisfies the constraints, and let $\bm U \bm \Gamma \bm U^H$ be its eigenvalue decomposition, where $\bm U = [\bm u_1\; \bm u_2\; \cdots\;\bm u_N]\in\mathbb{C}^{N\times N}$ is unitary and $\bm \Gamma =\diag (\gamma_1, \gamma_2, \ldots , \gamma_N)$, with $\gamma_1 \geq \gamma_2 \geq \cdots \geq \gamma_N \geq 0$. Let $\bm U^*\in\mathbb{C}^{N\times N}$ be any unitary matrix with $\frac{1}{\| \bm s \|} \bm s$ as its last column. Then, $\bm R_x' =\bm U^* \bm \Gamma (\bm U^*)^H$ still satisfies the power constraint and, by virtue of Lemma~\ref{SINR_bound}, $\SINR(\bm R_x',\bm s) \geq \SINR(\bm R_x, \bm s) \geq \rho_\text{min}$, so that also the SINR constraint is satisfied. The objective function is in turn increasing with $\bm s^H\big(\bm I_N +\frac{|h|^2}{\sigma^2_v} \bm R_x \big)^{-1} \bm s$, which, again based on Lemma~\ref{SINR_bound}, is upper-bounded as
\begin{align}
 \bm s^H\left(\bm I_N + \frac{|h|^2}{\sigma^2_v} \bm R_x \right)^{-1} \bm s  & \leq \frac{\| \bm s \|^2}{1+ \frac{|h|^2}{\sigma^2_v} \gamma_N}\notag\\
 & = \bm s^H\left(\bm I_N + \frac{|h|^2}{\sigma^2_v} \bm R_x' \right)^{-1} \bm s
\end{align}
whereby, $\CR(\bm R_x', \bm s) \geq \CR (\bm R_x,\bm s)$, and $\bm U^*$ is optimum. As to the eigenvalues, they must be determined by solving
\begin{equation}
 \begin{aligned}
 \max_{\gamma_1,\ldots,\gamma_N} &\quad \left\{ \sum_{i=1}^N \log \left(1+ \frac{|h|^2}{\sigma _v^2} \gamma_i \right)  + \beta \log \left( 1 + \frac{\frac{\sigma _f^2}{\sigma _v^2} \| \bm s \|^2}{1 +\frac{|h|^2}{\sigma _v^2} \gamma_N} \right)  \right\} \\
 \text{s.t.} &\quad \frac{\sigma^2_a \| \bm s \|^2}{\sigma^2_g \gamma_N +\sigma^2_c\| \bm s \|^2  + \sigma^2_w}\geq \rho_\text{min}\\
 & \quad \gamma_1 \geq \cdots \geq \gamma_N \geq 0 ,\quad \sum_{i=1}^N \gamma_i\leq NP_c.
 \end{aligned}
\end{equation}
For fixed $\gamma_N$, the objective function is maximized, due to Jensen's inequality~\cite{hazewinkel2013encyclopaedia}, by $\gamma_i = \frac{NP_c-\gamma_N}{N-1}$, $i = 1, \ldots,N-1$, leading to the optimization problem
\begin{equation}
 \begin{aligned}
 \max_{\gamma_N} &\quad  G'(\gamma_N) \\
 \text{s.t.} & \quad 0 \leq \gamma_N \leq \gamma'
\end{aligned}
\end{equation}
where
\begin{align}
 G'(\gamma_N)&=(N-1)\log \left(1+ \frac{|h|^2}{\sigma _v^2} \frac{NP_c - \gamma_N}{N-1} \right) \notag\\
 &\quad + (1-\beta) \log \left(1+ \frac{|h|^2}{\sigma _v^2} \gamma_N \right)\notag \\
&\quad + \beta \log \left( 1 + \frac{|h|^2}{\sigma_v^2} \gamma_N + \frac{\sigma _f^2 }{\sigma _v^2} \| \bm s \|^2\right)\\
\gamma' &= \min\left\{P_c, \frac{\sigma^2_a \| \bm s \|^2 -(\sigma^2_w+\sigma^2_c\| \bm s \|^2)\rho_\text{min}}{\sigma^2_g \rho_\text{min}}\right\}.
\end{align}
It can be easily checked that $G$ is continuously differentiable and convex, whereby the solution can either be a critical point of $G$ or a point of the boundary. Since the condition $\frac{d}{\gamma_N}G(\gamma_N)=0$ is equivalent to the quadratic equation $A\gamma_N^2 +B \gamma_N + C=0$, where
\begin{subequations}
\begin{align}
 A &=  - N |h|^4 \\
 B &= |h|^2 \left( N |h|^2 P_c - N \sigma_v^2 - (N-\beta) \sigma_f^2 \| \bm s \|^2  \right)\\
 C &=  |h|^2  N P_c \left( \sigma_v^2+ (1-\beta) \sigma_f^2 \| \bm s \|^2 \right) \notag\\
 &\quad  -\beta (N-1) \sigma_v^2 \sigma_f^2 \| \bm s \|^2
\end{align}%
\end{subequations}
the solution is
\begin{equation}\label{eq:optgammaN}
\gamma_N^* = \begin{cases}
0, & \text{if } \Delta \leq0 \text{ or} \\
& \Delta >0 \text{ and } \lambda_2\leq0 \text{ or}\\
& \Delta >0 \text{ and } \lambda_1\geq \gamma' \\
\lambda_2 , & \text{if } \Delta >0 \text{ and } 0< \lambda_2 < \gamma' \\
\argmax\limits_{\gamma_N\in\{0,\gamma'\}} G'(\gamma_N), & \text{otherwise}.
\end{cases}
\end{equation}
where $\Delta= B^2- 4 AC$ is the discriminant of the quadratic equation and $\lambda_1 < \lambda_2$ are its two real roots when $\Delta>0$. At this point $\gamma_i^* = \frac{NP_c - \gamma_N^*}{N-1}$ for $i = 1,2,\ldots,N-1$, and the optimal covariance is the one in~\eqref{opt_R_coher_fixed_s}.

\subsubsection{Joint optimization}\label{proof_coher_joint}

Suppose that the pair $(\bm R_x ,\bm s)$ satisfies the constraints; let $\bm U \bm \Gamma \bm U^H$ be the eigenvalue decomposition of $\bm R_x$, where $\bm U = [\bm u_1\; \bm u_2\; \cdots\;\bm u_N]\in\mathbb{C}^{N\times N}$ is unitary and $\bm \Gamma =\diag (\gamma_1, \gamma_2, \ldots , \gamma_N)$, with $\gamma_1 \geq \gamma_2 \geq \cdots \geq \gamma_N \geq 0$. Consider $\bm s'= \sqrt{\epsilon} \bm u_N$, where $\epsilon =\| \bm s \|^2$. Then, using the same arguments as those in the previous case, the pair $(\bm R_x,\bm s')$ still satisfies the constraints and $\CR(\bm R_x,\bm s')\geq \CR (\bm R_x, \bm s)$. Moreover, since the CR is easily seen to be independent of $\bm U$, the design reduces to
%\begin{equation}\label{eq:prob} one-column version
%\begin{aligned}
%\max_{\epsilon, \gamma_1,\ldots,\gamma_N} & \quad \left\{ \sum_{i=1}^N \log \left(1+ \frac{|h|^2}{\sigma _v^2} \gamma_i \right)  - \beta \log \left( \frac{1 + \frac{\sigma _f^2}{\sigma _v^2} \epsilon}{1 + \frac{\sigma _f^2 \epsilon}{\sigma _v^2} \left( 1 + \frac{|h|^2}{\sigma_v^2} \gamma_N \right)^{-1} } \right) \right\}\\
%\text{s.t.} & \quad 0 \leq \epsilon \leq N P_r, \quad \frac{\sigma^2_a \epsilon}{\sigma^2_g \gamma_N +\sigma^2_c \epsilon +\sigma^2_w} \geq \rho_\text{min}  \\
%& \quad 0 \leq \gamma_N \leq \cdots \leq \gamma_1 ,\quad \sum_{i=1}^N \gamma_i\leq NP_c
%\end{aligned}
%\end{equation}
\begin{equation}\label{eq:prob} % two-column version
\begin{aligned}
\max_{\epsilon, \gamma_1,\ldots,\gamma_N} & \quad \Bigg\{ \sum_{i=1}^N \log \left(1+ \frac{|h|^2}{\sigma _v^2} \gamma_i \right)  \\
&\quad \left. - \beta \log \left( \frac{1 + \frac{\sigma _f^2}{\sigma _v^2} \epsilon}{1 + \frac{\sigma _f^2 \epsilon}{\sigma _v^2} \left( 1 + \frac{|h|^2}{\sigma_v^2} \gamma_N \right)^{-1} } \right) \right\}\\
\text{s.t.} & \quad 0 \leq \epsilon \leq N P_r, \quad \frac{\sigma^2_a \epsilon}{\sigma^2_g \gamma_N +\sigma^2_c \epsilon +\sigma^2_w} \geq \rho_\text{min}  \\
& \gamma_1 \geq \cdots \geq \gamma_N \geq 0 ,\quad \sum_{i=1}^N \gamma_i\leq NP_c
\end{aligned}
\end{equation}
which admits a solution if condition~\eqref{existence_coher_joint} is satisfied. In this case, the constraints imply that $0\leq \gamma_N \leq \bar \gamma$, where
\begin{equation}
 \bar \gamma = \min \left\{ P_c,\frac{\sigma^2_a NP_r -(\sigma^2_w+\sigma^2_c NP_r)\rho_\text{min}}{\sigma^2_g \rho_\text{min}} \right\}. \label{gamma_bar_def}
\end{equation}
Now, for fixed $\gamma_N$, since the objective function is decreasing with $\epsilon$, we have that
\begin{equation}
 \epsilon=\frac{\rho_\text{min}(\sigma^2_g\gamma_N+\sigma^2_w)}{\sigma^2_a-\sigma^2_c\rho_\text{min}}.
\end{equation}
Furthermore, from Jensen's inequality, the objective function of Problem~\eqref{eq:prob} is maximized when $\gamma_i= \frac{NP_c - \gamma_N}{N-1}$, $i = 1,\ldots,N-1$. The problem therefore reduces to
\begin{equation}\label{eq:coherentG}
 \begin{aligned}
  \max_{\gamma_N} & \quad \bar G(\gamma_N) \\
\text{s.t.} & \quad 0 \leq \gamma_N \leq \bar \gamma
 \end{aligned}
\end{equation}
where
\begin{align}
\bar G(\gamma_N) &= (N - 1)\log \left( {1 + \frac{|h|^2}{\sigma _v^2}\frac{NP_c - {\gamma _N}}{N - 1}} \right)\notag\\
&\quad  + (1 - \beta )\log \left( {1 + \frac{|h|^2}{\sigma _v^2}{\gamma _N}} \right) \notag \\
&\quad - \beta \log \left( 1 + \frac{\sigma _f^2}{\sigma _v^2} \frac{\rho_\text{min}(\sigma^2_g\gamma_N+\sigma^2_w)}{\sigma^2_a-\sigma^2_c\rho_\text{min}} \right) \notag\\
&\quad + \beta \log \left( 1 + \frac{|h|^2}{\sigma _v^2}{\gamma _N} + \frac{\sigma _f^2}{\sigma _v^2}\frac{\rho_\text{min}(\sigma^2_g\gamma_N+\sigma^2_w)}{\sigma^2_a-\sigma^2_c\rho_\text{min}} \right). \label{function_G}
\end{align}
It can be easily seen that $\bar G$ is a continuously differentiable convex function, so that the solution is either a critical point of $\bar G$ or a boundary point. Moreover, the condition $\frac{d}{\gamma_N}G(\gamma_N)=0$ is equivalent to the cubic equation $ \bar A\gamma_N^3 +\bar B \gamma_N^2 + \bar C\gamma_N+\bar D=0$, where the expression of the coefficients is not reported for the sake of conciseness. Therefore, denoting $\bar \Delta$ the discriminant of the cubic equation, $\bar \lambda_0$ its unique real root, if $\bar \Delta>0$, or the largest of its two real roots, if $\bar \Delta =0$, and $\bar \lambda_1, \bar \lambda_2 , \bar \lambda_3$ its three real roots, if $\bar \Delta<0$, the solution is
\begin{equation}\label{opt_gamma_N_coher_joint}
 \gamma _N^* = \argmax_{\gamma_N\in \bar{\cal T}} \bar G(\gamma_N)
\end{equation}
where
%\begin{equation} \label{set_T} %one-column version
% \bar{\cal T} = \begin{cases}
% \left\{0, \bar \gamma, \bar \lambda_0 \mathbbm{1}_{\{0\leq \bar \lambda_0 \leq \bar \gamma \}}\right\}, & \text{if } \bar \Delta \geq 0\\[5pt]
% \left\{0, \bar \gamma, \bar \lambda_1 \mathbbm{1}_{\{0\leq \bar \lambda_1 \leq \bar \gamma \}}, \bar \lambda_2 \mathbbm{1}_{\{0\leq \bar \lambda_2 \leq \bar \gamma \}}, \bar \lambda_3 \mathbbm{1}_{\{0\leq \bar \lambda_3 \leq \bar \gamma \}}\right\},& \text{otherwise}
% \end{cases}
%\end{equation}
\begin{equation} \label{set_T} % two-column version
 \bar{\cal T} = \begin{cases}
 \left\{0, \bar \gamma, \bar \lambda_0 \mathbbm{1}_{\{0\leq \bar \lambda_0 \leq \bar \gamma \}}\right\}, & \text{if } \bar \Delta \geq 0\\[5pt]
 \Big\{ 0, \bar \gamma, \bar \lambda_1 \mathbbm{1}_{\{0\leq \bar \lambda_1 \leq \bar \gamma \}}, \\
 \bar \lambda_2 \mathbbm{1}_{\{0\leq \bar \lambda_2 \leq \bar \gamma \}}, \bar \lambda_3 \mathbbm{1}_{\{0\leq \bar \lambda_3 \leq \bar \gamma \}} \Big\}, & \text{otherwise}
 \end{cases}
\end{equation}
$\mathbbm{1}_{\cal A}$ denoting the indicator function of the event $\cal A$, i.e., $\mathbbm{1}_{\cal A}=1$, if $\cal A$ is verified, and $\mathbbm{1}_{\cal A}=0$, otherwise. At this point, the remaining eigenvalues are $\gamma_i^* = (NP_c - \gamma_N^*)/(N-1)$, while $\epsilon^* = \big(\rho_\text{min}(\sigma^2_g\gamma_N^*+\sigma^2_w)\big)/(\sigma^2_a-\sigma^2_c\rho_\text{min})$, so that the optimal covariance matrix and radar waveform are those in~\eqref{opt_R_coher_joint} and~\eqref{opt_s_coher_joint}, respectively.

\subsection{Solution to Problem~\eqref{opt_prob_incoher}} \label{proof_incoherent}

Suppose that $\bm R_x = \bm U \bm \Gamma \bm U^H$ and $\bm s$, with $\| \bm s \|^2=\epsilon$, satisfy the constraints. Then $\bm R_x' =\bm \Gamma$ and $\bm s' =(0\; \cdots \; 0\; \epsilon)$ still satisfy the power constraints and, by virtue of Lemma~\ref{SINR_bound}, $\SINR(\bm R_x',\bm s) \geq \SINR(\bm R_x, \bm s) \geq \rho_\text{min}$, so that also the SINR constraint is satisfied. Furthermore, $\CR (\bm R_x',\bm s') \geq \CR (\bm R_x, \bm s)$, since $R_0(\bm R_x)$ is independent of $\bm U$ and $\bm s$, while $R_1$, exploiting Hadamard's inequality, can be upper-bounded as follows
\begin{align}
 R_1 (\bm R_x, \bm s)&= \frac{1}{N}\log \det \Bigg( \bm I_N + \frac{|h|^2}{\sigma^2_v}\bm R_x \notag\\
 &\quad \left. \times \diag \left(\left\{ \left( 1 + \tfrac{\sigma_f^2}{\sigma^2_v} | s_i|^2\right)^{-1} \right\}_{i=1}^N\right) \right)\notag\\
 &\leq \frac{1}{N}\sum_{i=1}^N \log \left(1+ \frac{\frac{|h|^2}{\sigma^2_v}\gamma_i}{1+\frac{\sigma^2_f}{\sigma^2_v} s_i} \right)\notag\\
 & \leq \frac{1}{N}\sum_{i=1}^{N-1}\log  \left(1+ \frac{|h|^2}{\sigma^2_v} \gamma_i\right) \notag\\
 &\quad + \frac{1}{N}\log \left(1+ \frac{\frac{|h|^2}{\sigma^2_v}\gamma_N}{1+\frac{\sigma^2_f}{\sigma^2_v} \epsilon} \right)\notag\\
 &=R_1 (\bm R_x', \bm s')
\end{align}
which shows that the structure of $\bm s'$ and $\bm R_x'$ is optimal. As to the eigenvalues of the covariance matrix and the norm of the radar waveform, they can be determined by solving
%\begin{equation}\label{eq:P_joint} % one-column version
% \begin{aligned}
% \max_{\epsilon,\gamma_1,\ldots,\gamma_N} &\quad \left\{  \sum_{i=1}^N \log \left(1+ \frac{|h|^2}{\sigma _v^2} \gamma_i \right) -\beta\log \left( \frac{1+\frac{|h|^2}{\sigma _v^2} \gamma_N}{1+ \left(1+\frac{\sigma^2_f }{\sigma _v^2} \epsilon \right)^{-1} \frac{|h|^2}{\sigma _v^2} \gamma_N } \right) \right\} \\
%\text{s.t.} & \quad 0\leq \epsilon \leq N P_r, \quad \frac{\sigma^2_a \epsilon}{\sigma^2_g \gamma_N +\sigma^2_c \epsilon +\sigma^2_w} \geq \rho_\text{min}  \\
%& \quad 0 \leq \gamma_N \leq \cdots \leq \gamma_1 ,\quad \sum_{i=1}^N \gamma_i\leq NP_c
%\end{aligned}
%\end{equation}
\begin{equation}\label{eq:P_joint} % two-column version
 \begin{aligned}
 \max_{\epsilon,\gamma_1,\ldots,\gamma_N} &\quad \Bigg\{  \sum_{i=1}^N \log \left(1+ \frac{|h|^2}{\sigma _v^2} \gamma_i \right) \\
 &\quad \left.-\beta\log \left( \frac{1+\frac{|h|^2}{\sigma _v^2} \gamma_N}{1+ \left(1+\frac{\sigma^2_f }{\sigma _v^2} \epsilon \right)^{-1} \frac{|h|^2}{\sigma _v^2} \gamma_N } \right) \right\} \\
\text{s.t.} & \quad 0\leq \epsilon \leq N P_r, \quad \frac{\sigma^2_a \epsilon}{\sigma^2_g \gamma_N +\sigma^2_c \epsilon +\sigma^2_w} \geq \rho_\text{min}  \\
&  \gamma_1 \geq \cdots \geq \gamma_N \geq 0 ,\quad \sum_{i=1}^N \gamma_i\leq NP_c
\end{aligned}
\end{equation}
which is exactly Problem~\eqref{eq:prob}. Therefore, it admits a solution only if condition~\eqref{existence_incoher_joint} is satisfied, in which case
\begin{subequations}
 \begin{align}
 \gamma _N^* &= \argmax_{\gamma_N\in \bar{\cal T}} \bar G(\gamma_N)\label{opt_gamma_N_incoher_joint}\\
 \gamma_i& = \frac{NP_c - \gamma_N}{N-1}, \quad i = 1,\ldots,N-1\\
 \epsilon^*& =\frac{\rho_\text{min}(\sigma^2_g\gamma_N^*+\sigma^2_w)}{\sigma^2_a-\sigma^2_c\rho_\text{min}}
 \end{align}%
\end{subequations}
where $\bar G$ is the function in~\eqref{function_G} and $\bar{\cal T}$ is the set in~\eqref{set_T}, so that the optimal covariance matrix and radar waveform are those in~\eqref{opt_R_s_incoher_joint}.

\subsection{Solution to Problem~\eqref{opt_prob_colored}}\label{proof_colored}

Suppose that the pair $(\bm R_x ,\bm s$) satisfy the constraints. Let $\bm U \bm \Gamma \bm U^H$  be the eigenvalue decomposition of $\bm R_x$, where $\bm U = [\bm u_1\; \bm u_2\; \cdots\;\bm u_N]\in\mathbb{C}^{N\times N}$ is unitary and $\bm \Gamma =\diag (\gamma_1, \gamma_2, \ldots , \gamma_N)$, with $\gamma_1 \geq \gamma_2 \geq \cdots \geq \gamma_N \geq 0$; let also $\phi_N$ and $\bm v_N$ be the smallest eigenvalue of $\bm M$ and the corresponding eigenvector, respectively. Consider $\bm s'= \sqrt{\epsilon} \bm v_N$, where $\epsilon =\| \bm s \|^2$ and $\bm R_x' = \bm U^* \bm \Gamma (\bm U^*)^H$, where $\bm U^*\in\mathbb{C}^{N\times N}$ is any unitary matrix with $\bm v_N$ as last column. Then $(\bm R_x',\bm s')$ still satisfy the power constraints and, from Lemma~\ref{SINR_bound}, $\SINR(\bm R_x',\bm s) \geq \SINR(\bm R_x, \bm s) \geq \rho_\text{min}$, so that also the SINR constraint is satisfied. The objective function is in turn increasing with $\bm s^H (\bm I_N + \frac{|h|^2}{\sigma^2_v} \bm R_x )^{-1} \bm s$, which, again based on Lemma~\ref{SINR_bound}, is upper-bounded as
\begin{align}
 \bm s^H & \left(\bm I_N + \frac{|h|^2}{\sigma^2_v} \bm R_x \right)^{-1} \bm s \notag\\
 & = \bm s^H\left(\bm I_N + \frac{|h|^2}{\sigma^2_v} \bar{\bm U}\bar {\bm \Gamma} \bar{\bm U}^H + \frac{|h|^2}{\sigma^2_v}  \gamma_N\bm u_N \bm u_N^H \right)^{-1} \bm s \notag\\
 & \leq  \frac{\epsilon}{1 + \frac{\sigma^2_f}{\sigma^2_v} \gamma_N}\notag\\
 & = (\bm s')^H \left(\bm I_N + \frac{|h|^2}{\sigma^2_v} \bm R_x' \right)^{-1} \bm s' 
\end{align}
where $\bar{\bm U}=[ \bm u_1 \; \cdots \; \bm u_{N-1}]$ and $\bar{\bm \Gamma} =\diag \big(\{\gamma_i\}_{i=1}^{N-1}\big)$. This implies that $\CR(\bm R_x', \bm s') \geq \CR (\bm R_x,\bm s)$, and $\bm U^*$ is optimum. As to $\{\gamma_i\}_{i=1}^N$ and $\epsilon$, they can be determined by solving the optimization problem in~\eqref{eq:prob}, with $\sigma^2_w$ replaced by $\phi_N$. From Sec.~\ref{proof_coher_joint}, it admits a solution if~\eqref{existence_colored} is satisfied and
\begin{subequations}
\begin{align}
 \gamma _N^* & = \argmax_{\gamma_N\in \bar{\cal T}} \bar G(\gamma_N) \label{opt_gamma_N_colored}\\
 \gamma_i^* & = \frac{NP_c - \gamma_N^*}{N-1}, \quad i=1,\ldots, N-1\\
 \epsilon^* & =\frac{\rho_\text{min}(\sigma^2_g\gamma_N^*+\phi_N)}{\sigma^2_a-\sigma^2_c\rho_\text{min}}
\end{align}%
\end{subequations}
where $\bar G$ and $\bar{\cal T}$ are the function in~\eqref{function_G} and the set in~\eqref{set_T}, respectively, when $\sigma^2_w$ is replaced by $\phi_N$. The optimal covariance matrix and radar waveform are, therefore, those in~\eqref{opt_R_colored} and~\eqref{opt_s_colored}, respectively.

\subsection{Proof of Lemma~\ref{lemma_R0_R1_regions}}\label{proof_lemma_R0_R1_regions}

Let $\big(R_0',R_1'\big) \in {\cal S}(\sigma^2_w \bm I_N)$ be achievable with $\big(\bm R_x', \bm s'\big)$, and consider $\hat{\bm s}= \sqrt{\epsilon} \bm u_N$, where $\bm u_N$ is the eigenvector corresponding to the smallest eigenvalue of $\bm R_x'$ and $\epsilon = \| \bm s' \|^2$. Then, the point $\big(\bm R_x', \hat{\bm s}\big)$ satisfies all the constraints in~\eqref{achievable_region_expr}, including that on the SINR, since, from Lemma~\ref{SINR_bound},
\begin{align}
 \SINR \big(\bm R_x', \hat{\bm s}\big) & = \frac{\sigma^2_a \epsilon}{\sigma^2_g \gamma_N +\sigma^2_c \epsilon +\sigma^2_w}\notag\\
 & \geq \SINR \big(\bm R_x', \bm s' \big) \notag\\
 &\geq \rho_\text{min}.\label{eq_sinr_bound_1}
\end{align}
Therefore, letting $\hat R_1=R_1\big (\bm R_x', \hat{\bm s}\big)$, we have that $\big(R_0',\hat R_1\big) \in {\cal S}(\sigma^2_w \bm I_N)$. Furthermore, $\hat R_1 \geq R_1'$, since $R_1(\bm R_x, \bm s)$ is increasing with $\bm s^H \big( \bm I +\frac{|h|^2}{\sigma^2_v} \bm R_x \big)^{-1} \bm s$, and, exploiting again Lemma~\ref{SINR_bound},
\begin{align}
 (\bm s')^H  \left(\bm I_N +\frac{|h|^2}{\sigma^2_v} \bm R_x' \right)^{-1} \bm s' & \leq \frac{\epsilon}{1+ \frac{|h|^2}{\sigma^2_v} \gamma_N}\notag\\
&  =(\hat{\bm s})^H \left(\bm I_N +\frac{|h|^2}{\sigma^2_v} \bm R_x' \right)^{-1} \hat{\bm s} .
\end{align}

Let now $\phi_N$ be the smallest eigenvalue of $\bm M$ and $\bm v_N$ be the corresponding eigenvector, and consider $\bm s'' =\sqrt{\epsilon} \bm v_N$ and $\bm R_x''=  \bm U^* \bm \Gamma (\bm U^*)^H$, where $\bm U^*\in\mathbb{C}^{N\times N}$ is a unitary matrix whose last column is $\bm v_N$. Then, $\big( \bm R_x'', \bm s''\big)$ satisfies all the constraints in~\eqref{achievable_region_expr}, including the one on the SINR, since
\begin{align}
 \SINR \big(\bm R_x'', \bm s''\big)  & = \frac{\sigma^2_a \epsilon}{\sigma^2 \gamma_N +\sigma^2_c \epsilon +\phi_N} \notag\\
 & \geq \frac{\sigma^2_a \epsilon}{\sigma^2 \gamma_N +\sigma^2_c \epsilon +\sigma^2_w} \notag\\
 & \geq \rho_\text{min}
\end{align}
where the first inequality follows from the fact that $\gamma_N \leq \sigma^2_w$ and the second inequality from~\eqref{eq_sinr_bound_1}. Therefore, denoting $R_1''=R_1\big (\bm R_x'', \bm s''\big)$, we have that $\big(R_0',R_1''\big) \in {\cal S}(\bm M)$ and $R_1'' =\hat R_1 \geq R_1'$.

\end{document}